%% file: manuscript_v2.tex
\documentclass[%
preprint,
showpacs,
amsmath,amssymb,
aps,
pra,
]{revtex4-1}
\usepackage{amsmath,amsthm,amssymb,amsfonts,enumitem,float,hyperref}
\usepackage{graphicx}
\usepackage{dcolumn}
\usepackage{bm}
\usepackage{color}
\usepackage{verbatim}
\usepackage{url}
\hypersetup{
    colorlinks=true,       
    linkcolor=blue,          
    citecolor=blue,        
    filecolor=blue,      
    urlcolor=blue,           
    runcolor=blue
}

\begin{document}


\title{Quantum and classical annealing in a continuous space with multiple local minima}

\author{Yang Wei Koh}
\affiliation{Institute of Innovative Research, Tokyo Institute of Technology, Nagatsuta-cho, Midori-ku, Yokohama 226-8503, Japan}
\author{Hidetoshi Nishimori}

\affiliation{Institute of Innovative Research, Tokyo Institute of Technology, Nagatsuta-cho, Midori-ku, Yokohama 226-8503, Japan}
\affiliation{Graduate School of Information Sciences, Tohoku University, Sendai 980-8579, Japan}
\affiliation{RIKEN, Interdisciplinary Theoretical and Mathematical Sciences (iTHEMS),
Wako, Saitama 351-0198, Japan}

\date{\today}

\begin{abstract}
The protocol of quantum annealing is applied to an optimization problem with a one-dimensional continuous degree of freedom, a variant of the problem proposed by Shinomoto and Kabashima.  The energy landscape has a number of local minima, and the classical approach of simulated annealing is predicted to have a logarithmically slow convergence to the global minimum. We show by extensive numerical analyses that quantum annealing yields a power law convergence, thus an exponential improvement over simulated annealing.  The power is larger, and thus the convergence is faster, than a prediction by an existing phenomenological theory for this problem. Performance of simulated annealing is shown to be enhanced by introducing quasi-global searches across energy barriers, leading to a power-law convergence but with a smaller power than in the quantum case and thus a slower convergence classically even with quasi-global search processes.  We also reveal how diabatic quantum dynamics, quantum tunneling in particular, steers the systems toward the global minimum by a meticulous choice of annealing schedule. This latter result explicitly contrasts the role of tunneling in quantum annealing against the classical counterpart of stochastic optimization by simulated annealing. 

\end{abstract}
\maketitle


\section{Introduction}
\label{sec.Introduction}

Quantum annealing is a quantum-mechanical metaheuristic for optimization problems and has been applied predominantly to problems with discrete degrees of freedom, typically in terms of the transverse-field Ising model \cite{Kadowaki1998,Santoro2002,Das2008,Santoro:2006,Morita2008,Albash2018,Hauke2020,Farhi2001}. An early paper by Finnila {\em et al.} \cite{Finnila1994} discussed a system with continuous degrees of freedom but with an imaginary-time Schr\"odinger dynamics, thus short of intrinsic quantum effects.
Recent work by Abel and Spannowsky \cite{Abel2021PRXQ} applied the idea of domain-wall encoding \cite{Abel2021PRD} to approximately but accurately represent continuous degrees of freedom by Ising spins and illustrated quantum tunneling for a double-well potential by experiments on the D-Wave quantum annealing device \cite{Johnson2011}. Abel {\em et al.} \cite{Abel2021arxiv} extended their idea to more complex energy landscapes with several sharp local minima or a single deep global minimum and showed that quantum annealing as implemented on the D-Wave device achieves better performance than classical algorithms including simulated annealing. Most pertinent to the present paper is the contribution by Stella {\em et al.} \cite{Stella05}, in which they studied quantum annealing and classical simulated annealing for simple as well as non-trivial optimization problems in a continuous space. Their overall conclusion is that the quantum approach leads to a faster convergence to the global minimum than the classical one does. This result has been derived convincingly by numerical and analytical methods for the trivial case of harmonic oscillator as well as for the less trivial double-well potential. However, for the potential with very many local minima proposed and analyzed classically by Shinomoto and Kabashima \cite{Shinomoto91}, they adopted a phenomenological coarse-grained model, in which only discrete positions of local minima were considered and then a continuum limit to ignore the distance between local minima was taken to facilitate the analysis.  This procedure necessarily involves approximations, whose legitimacy is not necessarily evident in the long-time low-temperature limit in the classical case. This is true in the quantum case as well, and applicability of the approximation needs careful scrutiny.  Another important aspect is that the approximation makes it difficult to intuitively grasp the idea of how quantum dynamics promotes the flow of quantum-mechanical probability between adjacent local minima toward the goal of reaching the global minimum.

We are in this way motivated to study the problem of a Shinomoto-Kabashima-like potential with many local minima by direct numerical computations in the continuous space under classical as well as quantum annealing protocols.  Our results indicate that a flow of quantum-mechanical probability toward the global minimum enhances the performance of quantum annealing, in particular under meticulously designed non-linear annealing schedules. Also observed is an accelerated convergence of simulated annealing under quasi-global search processes across energy barriers in comparison with a conventional local search but still at a slower rate than in the quantum case.

The rest of the paper is organized as follows. In Sec. \ref{sec.potential}, we define the problem and give a discussion on the energy landscape of our potential. Section \ref{sec. Compare statics} examines the time-independent aspects of the system, particularly the quantum ground state and the classical equilibrium state. Sections \ref{sec. linear schedule} and \ref{sec. nonlinear schedule} are devoted to quantum annealing, with the former focusing on the linear annealing schedule and the latter on nonlinear ones for accelerated convergence. Sections \ref{sec.SA.log schedule} and \ref{sec.SA.linear schedules} are on simulated annealing. The former section focuses on the linear schedule, while the latter section deals with the logarithmic schedule considered by Shinomoto and Kabashima. Lastly in Sec. \ref{sec.Summary and discussions}, we summarize the paper and then conclude with some discussions. Additional information is provided in Appendixes.


\section{The problem}
\label{sec.potential}

In this section, we first define the problem and show its salient features.
\subsection{Definition of the problem}
\begin{figure}[h]
\begin{center}
\includegraphics[scale=1.0]{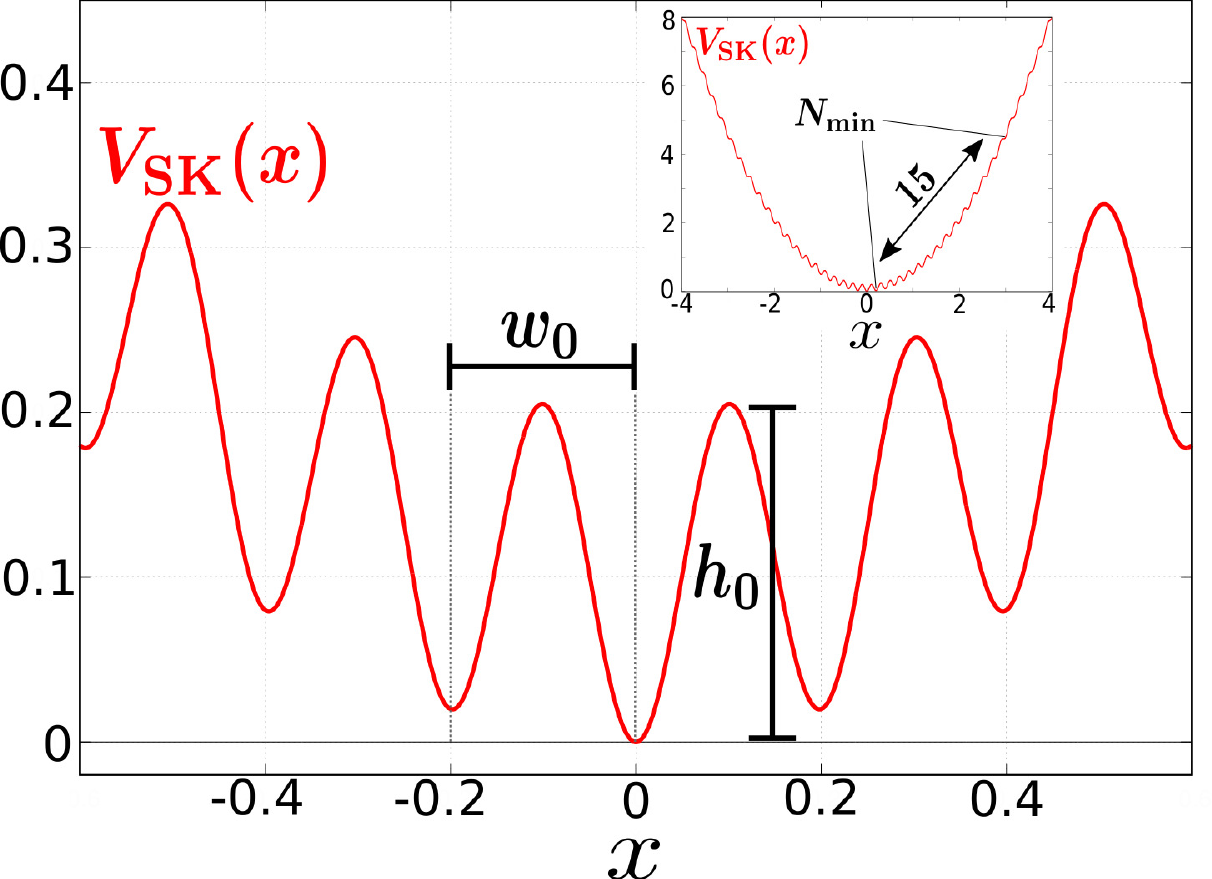}
\caption{Graph of the potential $V_{\rm SK}(x)$ in the vicinity of the global minimum at $x=0$, with parameters $h_0=w_0=0.2$. These parameters correspond, around the global minimum, approximately to the height of the energy barrier and the distance between neighboring minima. Inset: Zoomed-out view of the potential with the same parameter values, showing the 15 local minima (denoted $N_{\rm min}$) on each side of the origin. Due to the harmonic term in Eq. (\ref{eq.VSK(x) definition}), the local minima are smoothed out far away from the origin.}
\label{fig.VSK(x)}
\end{center}
\end{figure}

Let us consider the following potential inspired by Shinomoto and Kabashima \cite{Shinomoto91},
\begin{equation}
V_{\rm SK}(x)=\frac{1}{2} k x^2 + \frac{h_0}{2} \left[ 1 - \cos\left(\frac{2 \pi x}{w_0}\right) \right].
\label{eq.VSK(x) definition}
\end{equation}
The first is a harmonic term forming the envelope of the overall potential, and it sets a unique global minimum at $x=0$. Without loss of generality, we shall let the spring constant $k$ be unity from now on. The cosine term introduces local minima onto the energy landscape, and the structure of these minima can be tuned using two parameters $h_0$ and $w_0$. As an example, Fig. \ref{fig.VSK(x)} shows the graph of $V_{\rm SK}(x)$ when $h_0=w_0=0.2$. The parameter $h_0$ corresponds approximately to the height of the energy barrier between two adjacent minima, and $w_0$ to the horizontal distance between them.
Our proposed potential $V_{\rm SK}(x)$ tries to capture the essential physics underlying Shinomoto and Kabashima's idea. Nevertheless, there are some differences because in their original formulation, it is assumed that the local minima are all uniform such that the barrier height and distance between neighboring minima remain constant ad infinitum along the $x$-axis. On the other hand, in Eq. (\ref{eq.VSK(x) definition}), due to the presence of the quadratic term, the local minima of $V_{\rm SK}(x)$ are smoothed out far away from the origin and so the total number of local minima is in fact finite. This is illustrated by Fig. \ref{fig.VSK(x)}, where the inset shows the global view of the same graph. Notwithstanding this difference, Eq.~(\ref{eq.VSK(x) definition}) is expected to lead to essentially the same asymptotic behavior of the system because the probability of the system being far from the origin is very small in the long-time limit. We adopt Eq.~(\ref{eq.VSK(x) definition}) because of its simpler numerical implementation and yet with essentially the same asymptotic properties to be expected.

\subsection{Number of local minima}
Let us define $N_{\rm min}$ as the number of local minima in the positive $x$-direction. It is seen that $N_{\rm min}=15$ when the parameters are $h_0=w_0=0.2$. Notice that although the local minima are relatively uniform in the vicinity of the origin, they become shallower further away, so our studies should be focused on the former region in order to stay faithful to Shinomoto and Kabashima's proposal.

\begin{figure}[h]
\begin{center}
\includegraphics[scale=0.5]{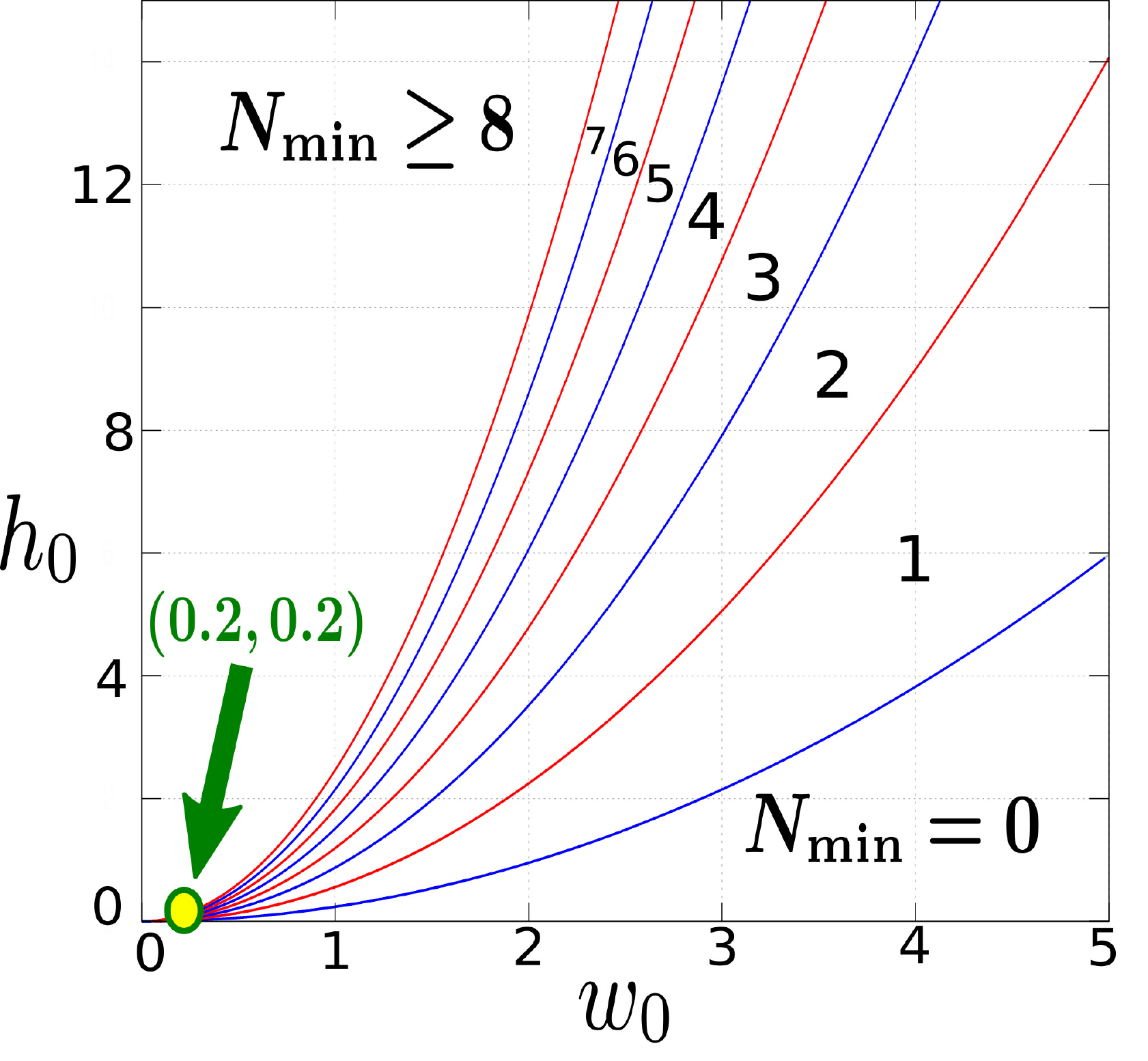}
\caption{Phase diagram showing $N_{\rm min}$, the number of local minima of $V_{\rm SK}(x)$, as a function of the parameters $h_0$ and $w_0$. Individual phases from $N_{\rm min}=0$ up to 7 are shown. In the region $N_{\rm min}\ge 8$, the phase boundaries become increasingly closely spaced, with $N_{\rm min}$ approaching infinity as one approaches the vertical axis $w_0=0$. This paper focuses on the point $h_0=w_0=0.2$ indicated by the circle, which lies within the region $N_{\rm min}\ge 8$.}
\label{fig.N_min}
\end{center}
\end{figure}

We calculated $N_{\rm min}$ numerically as a function of the parameters $h_0$ and $w_0$, and the results are shown in Fig. \ref{fig.N_min} in the form of a phase diagram. Individual regions of $N_{\rm min}$ from 0 up to 7 are shown. In the region $N_{\rm min}\ge 8$, the boundaries between successive regions become increasingly closely spaced, with $N_{\rm min}$ approaching infinity as one approaches the vertical axis. The graph of Fig. \ref{fig.VSK(x)} is indicated by the circle, located inside the region $N_{\rm min}\ge 8$. Roughly speaking, one can interpret the phase diagram of $N_{\rm min}$ as a map of the difficulty of optimizing the potential $V_{\rm SK}(x)$. Regions where $N_{\rm min}$ is large are, intuitively, more difficult to optimize than regions where it is smaller. Within a region of constant $N_{\rm min}$, it is generally more difficult to optimize if either (or both) of the parameters $h_0$ and $w_0$ increases.

It is not feasible to perform a thorough computational study of the entire phase space. Instead, we shall focus on just one point, $(w_0,h_0)=(0.2,0.2)$, which has been introduced in Figs. \ref{fig.VSK(x)} and \ref{fig.N_min}. Our choice is due to the need to balance two competing factors. On the one hand, the energy landscape of the potential should be difficult enough to be non-trivial for our annealing algorithms to optimize. On the other hand, we are limited by the computational resources needed to solve the time-dependent Schr\"{o}dinger equation. For instance, the top left corner of the phase space where $N_{\rm min}$ and $h_0$ are large would pose a good challenge to both quantum and simulated annealing; however, a very large number of grid points is required to accurately represent the wavefunction in this region of phase space, making it prohibitively expensive to compute energy eigenvalues and perform long annealing simulations. The energy landscape at the phase point $(0.2,0.2)$ is complex and non-trivial, but at the same time also computationally feasible to simulate, allowing us to focus on the physics of annealing rather than on the optimization of a potential with true technical difficulties. Hence, in the rest of the paper, unless otherwise stated, it is implicit that all numerical calculations are performed with the parameters $h_0=w_0=0.2$.


\section{Comparison of quantum ground state and classical equilibrium state}
\label{sec. Compare statics}

Before discussing annealing, which involves dynamics, let us first examine the time-independent aspects of the model. The Hamiltonian, applicable to both the quantum and classical descriptions of the system, is given by
\begin{equation}
H=\frac{p^2}{2m} + V_{\rm SK}(x),
\label{eq. Hamiltonian definition }
\end{equation}
where $p$ and $m$ are the momentum and mass of the particle, respectively. In the quantum description, $p$ is replaced by the momentum operator $\frac{\hbar}{i}\frac{\partial}{\partial x}$ ($\hbar=1$) and $x$ by the position operator. The Hamiltonian operator is then parameterized by the mass $m$ which also serves as the annealing parameter during optimization.

In the classical case, the system is treated using statistical mechanics where it is described by the partition function. The kinetic part of Eq. (\ref{eq. Hamiltonian definition }) can be integrated out analytically, and we only need to work with the position space Boltzmann distribution given by 
\begin{equation}
\varrho(x)=\frac{e^{-\beta V_{\rm SK}(x)}}{Z(\beta)},
\label{eq.Boltzmann distribution definition}
\end{equation}
where $\beta$ is the inverse temperature ($k_{\rm B}=1$), and $Z(\beta)$ is the partition function.
We shall mostly be concerned with the internal energy (i.e. the average energy) of the system
\begin{equation}
U(\beta)=\frac{1}{2\beta} + \langle V_{\rm SK}(x) \rangle_{\beta},
\label{eq.U definition}
\end{equation}
where the first term is the contribution from the kinetic energy ($m=1$) and the second one is the average potential energy
\begin{equation}
\langle V_{\rm SK}(x) \rangle_{\beta}= \int_{-\infty}^{\infty} V_{\rm SK}(x) \, \varrho(x) \, dx.
\label{eq.average potential energy.definition}
\end{equation}
The subscript $\beta$ in $\langle\cdot\rangle_{\beta}$ is to indicate the dependence on the inverse temperature via the Boltzmann distribution $\varrho(x)$. Thus, in contrast to the quantum scenario, we see that the classical energy is parameterized by $\beta$ during annealing.

\begin{figure}[h]
\begin{center}
\includegraphics[scale=0.65]{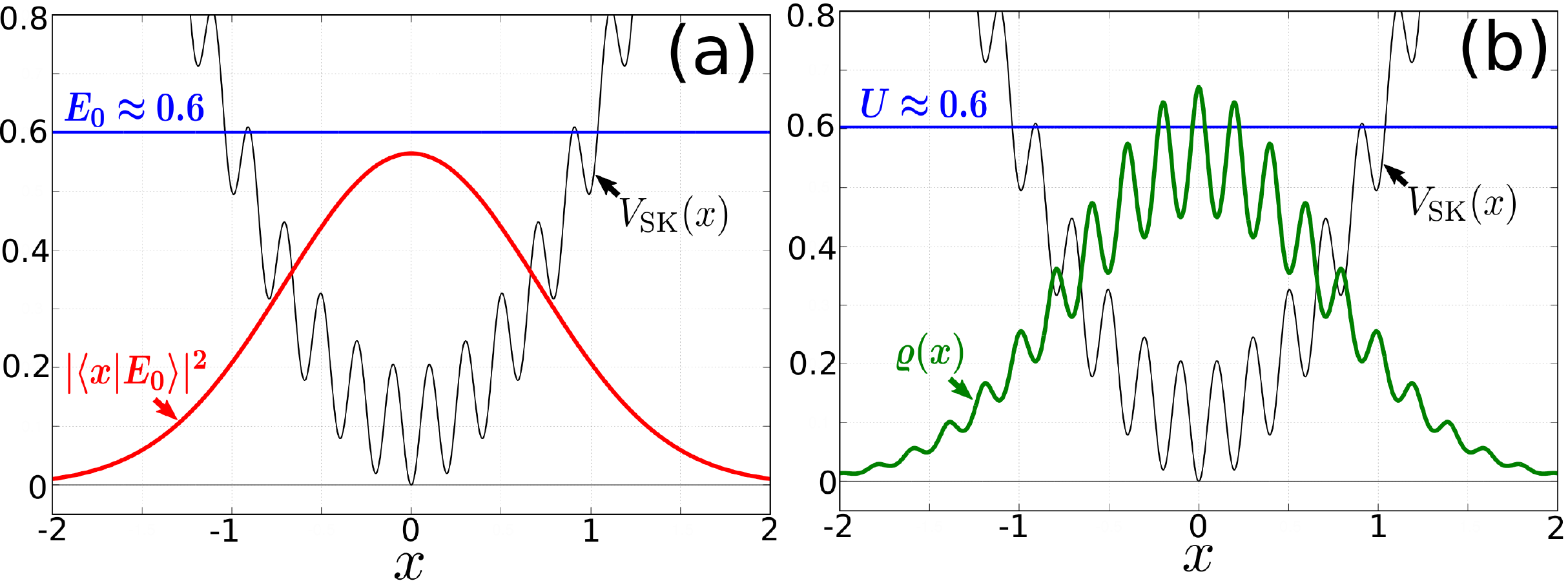}
\caption{Comparing the quantum ground state and classical equilibrium state of $V_{\rm SK}(x)$ with the same energies. The vertical axes reflect the actual values of all quantities plotted. (a) Quantum. The ground state probability density (red) with energy eigenvalue $E_0\approx 0.6$ (blue).  (b) Classical. The Boltzmann distribution (green) with internal energy $U\approx 0.6$ (blue). The precise energies and their corresponding parameters ($m$ and $\beta$) are summarized in Table \ref{tab.summary of boundaries of annealing stages} (labelled $a$).}
\label{fig. compare quantum classical groundstates . 0.6}
\end{center}
\end{figure}

As energy is commonly used as a gauge to monitor the progress of an annealing process, let us compare the quantum ground state and classical equilibrium state at the same energies. Figure \ref{fig. compare quantum classical groundstates . 0.6} shows the situation when the energies are approximately 0.6. Panel (a) shows the quantum case where the ground state energy eigenvalue $E_0$ is plotted in blue and its corresponding probability density $|\langle x| E_0\rangle|^2$ (i.e. the ground state wavefunction squared) is plotted in red. All quantities are plotted to scale and the vertical axis reflects their actual numerical values. Hence, by comparing the blue line to the potential $V_{\rm SK}(x)$ (black curve), one can say that the energy is high because many local minima are `submerged' below the blue line. The quantum ground state is delocalized in the sense that it spans across several local minima. Panel (b) depicts the classical situation, where the internal energy $U$ is plotted in blue and the Boltzmann distribution $\varrho(x)$ [Eq. (\ref{eq.Boltzmann distribution definition})] in green. Noteworthy is the fact that, apart from some wave-like oscillations, the classical equilibrium state is also delocalized, spanning local minima, in a way similar to its quantum counterpart. This means that when the energy is high, there is not much quantitative difference between the quantum ground state and classical equilibrium state for exploring the configuration space of the potential $V_{\rm SK}(x)$.

\begin{figure}[h]
\begin{center}
\includegraphics[scale=0.65]{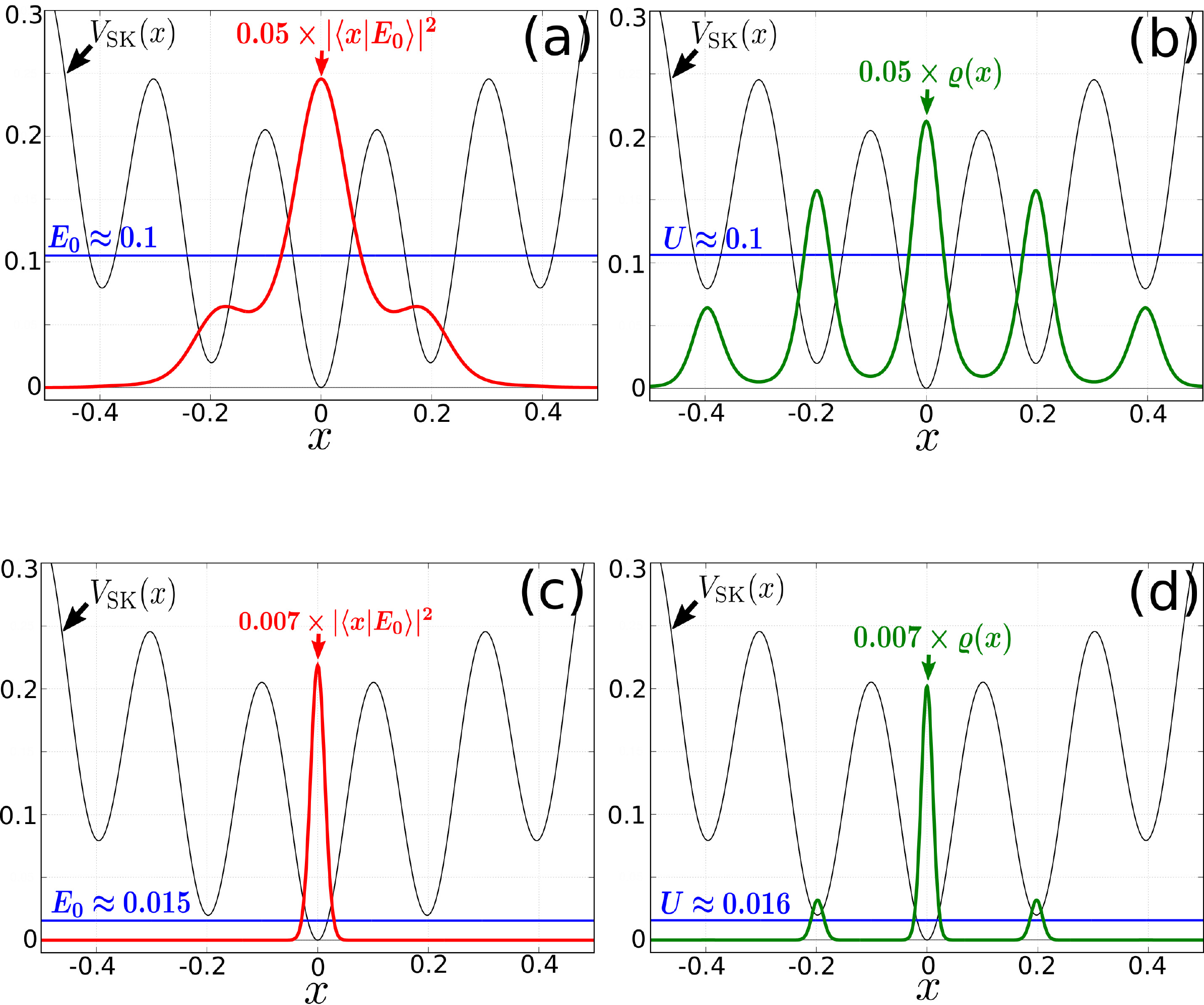}
\caption{Comparing the quantum ground state and classical equilibrium state at lower energies. The panels are organized similar to Fig. \ref{fig. compare quantum classical groundstates . 0.6}. Panels (a) and (b): Comparison at energy $\approx 0.1$. The precise energies and corresponding $m$ and $\beta$ are listed in Table \ref{tab.summary of boundaries of annealing stages}, labelled $b$. Panels (c) and (d): At energy $\approx 0.01$, and labelled $c$ in Table \ref{tab.summary of boundaries of annealing stages}. Note that the states have been rescaled for presentation purposes.}
\label{fig. compare quantum classical groundstates . 0.1 and 0.01}
\end{center}
\end{figure}

This analogous behavior of the two states, however, undergoes a dramatic change when the energy is lowered. Figure \ref{fig. compare quantum classical groundstates . 0.1 and 0.01} compares them again at energies around 0.1 [panels (a) and (b)] and at around 0.01 [panels (c) and (d)]. At 0.1, one sees that the quantum ground state has acquired a compact form concentrated around the global minimum. On the other hand, the classical equilibrium state fragments into a multi-modal distribution, with significant probability of finding the particle even at local minima which are relatively far away from the origin. At the low energy of 0.01, the difference between quantum and classical is even more pronounced, with the Boltzmann distribution manifesting two `spurious' modes. These spurious modes are metastable (i.e. long-lived) as the probability channel connecting them to the main mode at the origin is almost completely suppressed.

Based on these comparisons, we see that from an energy viewpoint the quantum ground state appears to be more suitable for the task of optimization than the classical one. At low energies, the former is more focused around the global minimum, whereas the latter exhibits signs of trappings by local minima. As a heuristic explanation, one might say that classical mechanics penalizes the particle in regions where its total energy is smaller than the potential energy, and so the equilibrium state tends to distribute itself around local minima to avoid incurring this penalty. On the other hand, quantum mechanics allows the particle to access classically forbidden regions via tunneling, and this freedom enables its ground state to achieve a more compact form concentrated around the global minimum.

\begin{figure}[h]
\begin{center}
\includegraphics[scale=0.65]{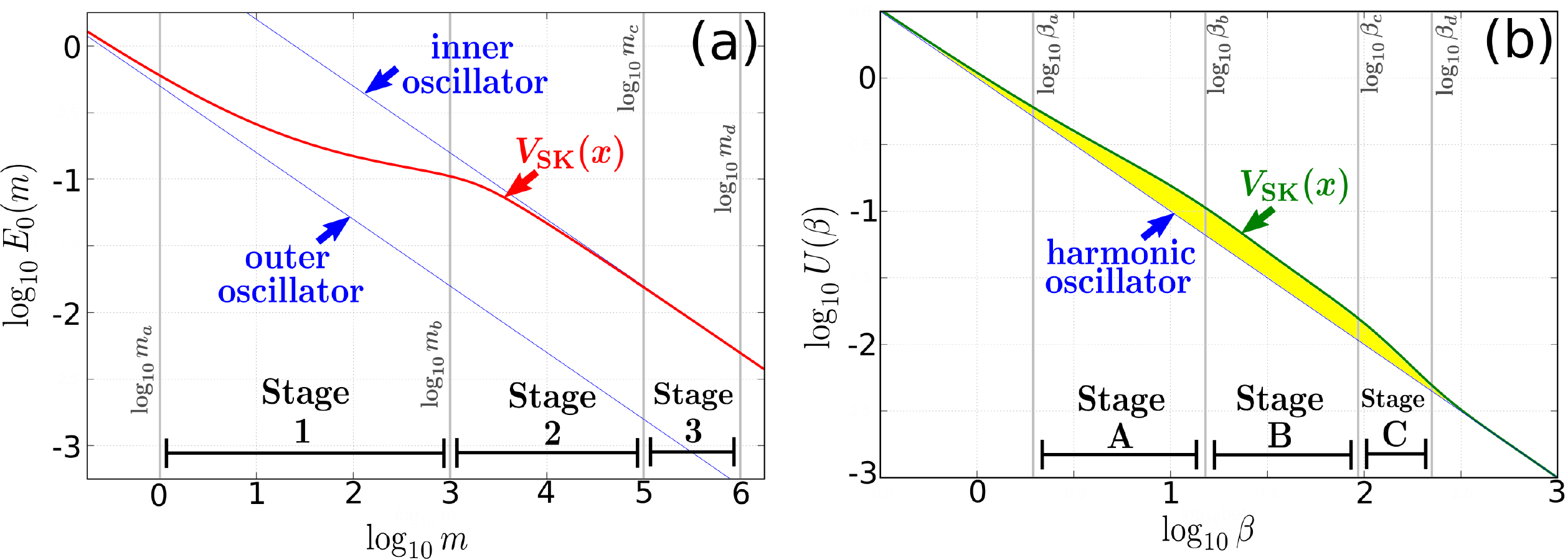}
\caption{The quantum ground state and classical equilibrium state energies of $V_{\rm SK}(x)$ as a function of their respective parameters. (a) The red curve shows the quantum energy eigenvalue $E_0$ as a function of mass $m$. The lower blue line shows the zero-point energy of a harmonic oscillator with $k=1$, and the upper line shows one with $k=99.696$. (b) The classical internal energy $U$ as a function of inverse temperature $\beta$ (green), and the equipartition theorem for the harmonic oscillator (blue). The region between the two curves (shaded yellow) shows the `excess energy' of the system. In both panels, the vertical lines indicate the boundary (i.e. initial and final) parameter values of the various annealing stages. These parameters are summarized in Table \ref{tab.summary of boundaries of annealing stages}. Stages are defined in Secs. \ref{sec. linear schedule} and \ref{sec.SA.linear schedules}.}
\label{fig. E0 and U vs m and beta}
\end{center}
\end{figure}

Let us now look at how the quantum and classical energies vary with mass and temperature. Figure \ref{fig. E0 and U vs m and beta}(a) shows the quantum ground state energy $E_0$ as a function of mass $m$ (red). To discern the essential features of $V_{\rm SK}(x)$, the zero-point energy, $\frac{1}{2}\sqrt{k/m}$, of two harmonic oscillators labelled `outer' ($k=1$) and `inner' ($k=99.696$) are also shown (blue lines). The outer oscillator describes the envelope of $V_{\rm SK}(x)$, while the inner oscillator is a quadratic approximation of the potential around the global minimum. In the limits of small and large masses, the ground state energy is well-described by the outer and inner oscillators. The range of $m$ where $E_0$ makes a transition between the two oscillators marks the important region of the quantum system.

The classical case is shown in panel (b) where the internal energy $U$ is plotted as a function of the inverse temperature $\beta$ (green). The internal energy of the harmonic oscillator, independent of spring constant as embodied by the equipartition theorem, is also shown (blue). Analogous to the quantum case, in the limit of small and large $\beta$, the classical ground state energy again approaches that of the oscillator. On the other hand, the distinctive feature here is that the non-trivial aspect of $V_{\rm SK}(x)$ expresses itself in the form of the shaded region (yellow), where the system seems to have `excess energy' relative to the oscillator. The range of $\beta$ where $U$ contains significant excess energy defines the crucial region for the classical system.

\begin{figure}[h]
\begin{center}
\includegraphics[scale=0.65]{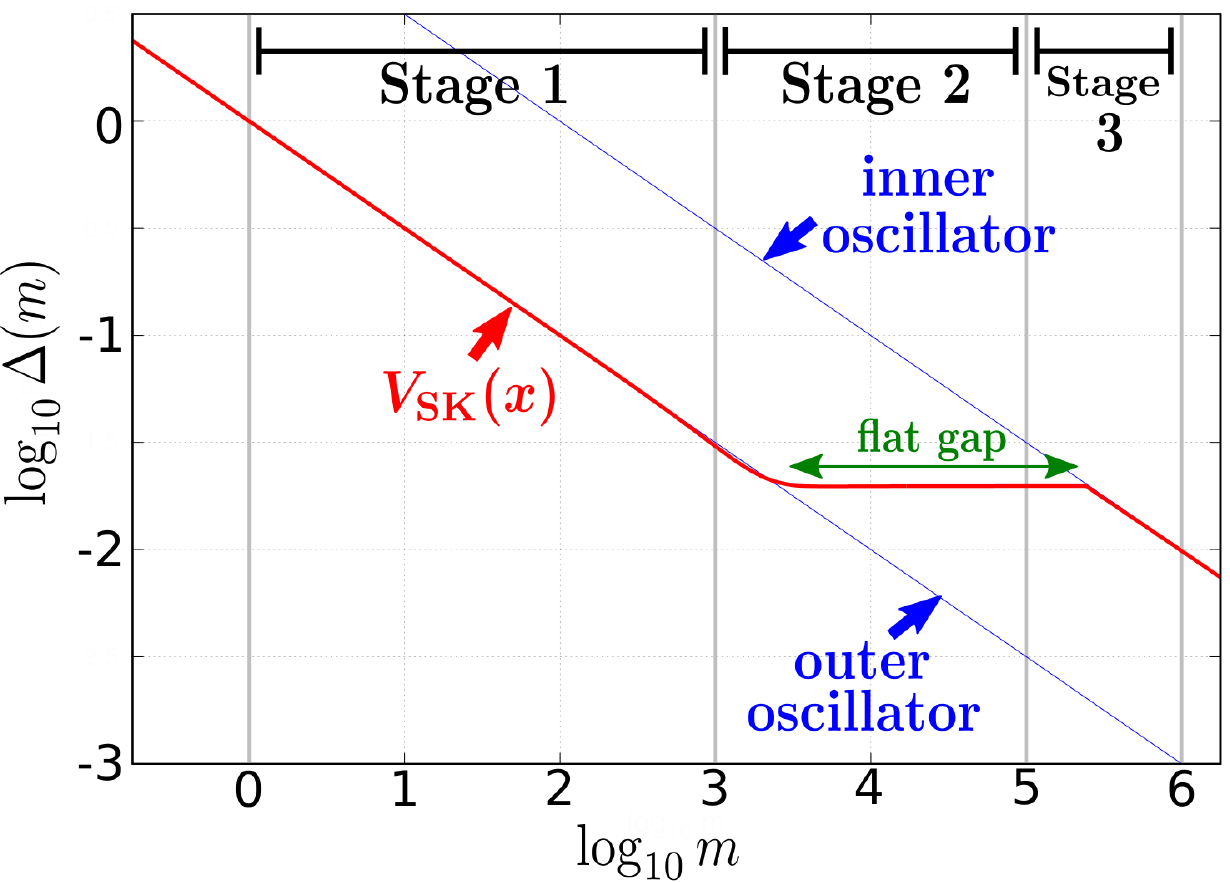}
\caption{The quantum energy gap $\Delta$ as a function of mass $m$. The red curve shows that of $V_{\rm SK}(x)$, and the blue lines show those of the oscillators previously discussed in Fig. \ref{fig. E0 and U vs m and beta}(a). The three quantum annealing stages are also shown. Stages 1, 2, and 3 are defined in Sec.~\ref{sec. linear schedule}.}
\label{fig. gap vs m }
\end{center}
\end{figure}

Another important aspect of the system is the quantum energy gap $\Delta$, defined as the energy difference between the ground and first excited states $E_1-E_0$. In quantum annealing, fast closure of the energy gap is a sign of first-order phase transition and forebodes long annealing times \cite{Albash2018}. Figure \ref{fig. gap vs m } shows the gap $\Delta$ of the potential $V_{\rm SK}(x)$ as a function of $m$ (red). Once again, we show the gaps of the outer and inner oscillators (given by $\sqrt{k/m}$ and plotted in blue) to highlight those features which are particular to $V_{\rm SK}(x)$. We see that the gap of $V_{\rm SK}(x)$ does not exhibit any anomalously small behavior since its magnitude lies mostly (although not strictly) between those of the two oscillators.  This suggests a polynomial computational complexity for quantum annealing, which will be confirmed numerically.  Therefore this is not a hard problem to solve from the viewpoint of computational complexity, but our point is to compare the performance of quantum and classical protocols, not to discuss difficulties within the framework of quantum annealing {\em per se}.

Apart from the absence of closure, let us also point out the segment of the gap curve indicated as `flat gap'. The gradient $\frac{\partial \Delta}{\partial m}$ stays zero throughout the segment, with the first excited state being multiply degenerate. This feature is not peculiar to the phase point $h_0=w_0=0.2$, and is in fact a persistent motif of the gap curve throughout the $h_0$-$w_0$ phase space. We shall elaborate more on this in Appendix \ref{app. overview of E0 and gap in phase space}. The implications of this `flat gap' for quantum annealing is not immediately clear, but we shall see later that the performance seems to be slightly lower here compared to other regions where it is absent. 

Lastly, let us remark that the ground state properties at the phase point $(w_0,h_0)=(0.2,0.2)$ presented in this section can be considered a good representative of the general behavior in phase space. Nevertheless, we also found  certain aspects which are not captured by $(0.2,0.2)$. To supplement the discussion given here, we briefly touch upon the quantum energy $E_0$ and the gap $\Delta$ of other phase points in Appendix \ref{app. overview of E0 and gap in phase space}.


\section{Quantum annealing using a linear schedule}
\label{sec. linear schedule}
We now analyze the dynamics, first for the quantum case under a simple linear annealing schedule.
\subsection{Linear schedule}

Quantum annealing of the potential $V_{\rm SK}(x)$ is performed by solving the time-dependent Schr\"{o}dinger equation
\begin{equation}
\left[
i\hbar\frac{\partial}{\partial t}
+
\frac{\hbar^2}{2m(t)}\frac{\partial^2}{\partial x^2}
-
V_{\rm SK}(x)
\right]
\psi(x,t)
=0,
\label{eq. TDSE }
\end{equation}
where $\psi(x,t)$ is the wavefunction, and the time dependence of mass $m(t)$ serves as the annealing schedule. Conceptually speaking, one starts annealing from zero mass where the kinetic energy is dominating, and slowly increase it to infinity such that the potential $V_{\rm SK}(x)$ becomes the dominating term. If the change of $m(t)$ with time is infinitesimally slow, then according to the adiabatic theorem \cite{Albash2018}, a wavefunction initially at the ground state will evolve into the global optimized solution of $V_{\rm SK}(x)$ upon the completion of annealing.

 Let us restrict the time evolution under Eq. (\ref{eq. TDSE }) to within the time interval $0 \le t \le T$ where $T$ is the total annealing time. Consider the following linear form for $m(t)$
\begin{equation}
m^{(1)}(t)=\left( m_f- m_i \right)\left(\frac{t}{T}\right) + m_i~,
\label{eq.linear schedule definition}
\end{equation}
where $m_i$ and $m_f$ are the masses at the initial $(t=0)$ and final $(t=T)$ times, respectively. When $m_i$ and $m_f$ are fixed, $T$ determines the annealing speed, with a large $T$ giving slow annealing. The superscript (1) denotes `first-order', and is to differentiate Eq. (\ref{eq.linear schedule definition}) from higher-order, nonlinear schedules which we shall introduced later.
As suggested by Morita \cite{Morita07}, only the initial and final time derivatives of annealing schedule determine the final excitation probability for sufficiently large annealing time.  It is thus expected that an explicit form of the annealing function is unimportant, i.e. it is irrelevant whether we change the mass as in Eq.~(\ref{eq.linear schedule definition}) or change the coefficient $\Gamma(t)\equiv\hbar/2m$ of the kinetic energy term as a whole as was done in Ref.~\cite{Stella05}, as long as the initial and final time derivatives are shared among different functions. The linear schedule of Eq.~(\ref{eq.linear schedule definition}) represents the case with finite derivatives at both ends of annealing.

\subsection{Stages of annealing}

One operational constraint faced when quantum annealing is implemented numerically via the solution of Eq. (\ref{eq. TDSE }) concerns the choice of the mass parameters $m_i$ and $m_f$. In principle, one should be able to use arbitrary small $m_i$ and arbitrary large $m_f$ in the schedule Eq. (\ref{eq.linear schedule definition}). For instance, one would like to start from the high energy state shown in Fig. \ref{fig. compare quantum classical groundstates . 0.6}(a) and anneal towards the low energy state shown in Fig. \ref{fig. compare quantum classical groundstates . 0.1 and 0.01}(c). In practice, however, it is not feasible to solve Eq. (\ref{eq. TDSE }) numerically with such a large energy difference because the number of grid points needed to represent the wavefunction is very large. In particular, one sees from Figs. \ref{fig. compare quantum classical groundstates . 0.6}(a) and \ref{fig. compare quantum classical groundstates . 0.1 and 0.01}(c) that the wavefunction evolves from a delocalized to a localized form, and so the grid must be wide enough to represent the initial wavefunction, while at the same time its points must also be dense enough to resolve the final wavefunction. Hence, in practice, the choice of $m_i$  and $m_f$ is partly constrained by the computational resources, both in terms of memory and processing speed, one has at one's disposal. In all our quantum annealing simulations, we restricted ourselves to a plane wave grid with 2048 grid points, and then choose $m_i$ and $m_f$ that are computationally feasible with this grid size.

Despite the above issue, our ultimate goal is still to attain some understanding of annealing when the initial mass is small and the final mass is large. Consider the masses $m_a$ and $m_d$ listed in Table \ref{tab.summary of boundaries of annealing stages}, where the difference $m_d-m_a$ spans six orders of magnitude. The ground state at $m_a$ was discussed in Fig. \ref{fig. compare quantum classical groundstates . 0.6}(a). To get an intuitive sense of the ground state at $m_d$, note that $m_d$ is larger than $m_c$, whose ground state was discussed in Fig. \ref{fig. compare quantum classical groundstates . 0.1 and 0.01}(c). Our objective is to study quantum annealing from the high energy, delocalized state at $m_a$ to the low energy, localized state at $m_d$.

To traverse this large mass difference, we perform annealing in stages. Referring to Table \ref{tab.summary of boundaries of annealing stages}, let us define the annealing from $m_a$ to $m_b$ as Stage 1, that from $m_b$ to $m_c$ as Stage 2, and that from $m_c$ to $m_d$ as Stage 3. The properties of the ground states at $m_a$, $m_b$, and $m_c$ have been discussed in Sec. \ref{sec. Compare statics}. Apart from keeping the computations manageable, the stages are also set up based on energy considerations. In Stage 1, the energy decreases from a high level to half the barrier height of $V_{\rm SK}(x)$, as indicated by the blue lines in Figs. \ref{fig. compare quantum classical groundstates . 0.6}(a) and \ref{fig. compare quantum classical groundstates . 0.1 and 0.01}(a). In Stage 2, the energy continues its descent to the potential level of the two lowest local minima, as seen in Fig.  \ref{fig. compare quantum classical groundstates . 0.1 and 0.01}(c). Finally, in Stage 3 the energy descends to close to the potential of the global minimum. 

A graphical perspective of the annealing stages is also given in Figs. \ref{fig. E0 and U vs m and beta}(a) and \ref{fig. gap vs m } where they are superposed upon the $E_0$ and $\Delta$ curves. In these figures, the vertical lines indicate the initial and final masses of each of the stages. 

\begin{table}[h]
\begin{center}
\begin{tabular}{c c cc c cc}
\hline
\hline
 &  & \multicolumn{2}{c}{Quantum Annealing} & \hspace{0.5cm} & \multicolumn{2}{c}{Simulated Annealing} \\
\cline{3-4} \cline{6-7} 
$l$ (label) &\hspace{0.25cm} & $\log_{10}m_l$ & $E_0(m_l)$ & \hspace{0.25cm} &  $\log_{10}\beta_l$ & $U(\beta_l)$ \\
\cline{1-1} \cline{3-4} \cline{6-7} 
$a$ & & 0 & 0.5999898 & & 0.29 & 0.6031582 \\
$b$ & & 3 & 0.1050870 & & 1.18 & 0.1061226 \\
$c$ & & 5 & 0.0154758 & & 1.97 & 0.0156931 \\
$d$ & & 6 & 0.0049617 & & 2.35 & 0.0049526 \\
\hline
\hline
\end{tabular}
\caption{The boundary parameters of the various annealing stages, as well as their corresponding energies. Each $m$ and $\beta$ is labelled, and their roles as the initial and/or final parameter of the stages are shown in Fig. \ref{fig. E0 and U vs m and beta}.}
\label{tab.summary of boundaries of annealing stages}
\end{center}
\end{table}

In numerical calculations, the grid in each stage is determined by making sure that it is wide enough to accommodate the ground state wavefunction at $m_i$ and its grid points dense enough to resolve the ground state wavefunction at $m_f$. This is done quantitatively by making sure that the eigenenergy $E_0$ is converged with respect to the number of grid points at both $m_i$ and $m_f$ 
\footnote{More concretely, we first determine the limit $x_{\rm max}$ such that the ground state wavefunction at $m_i$ lies comfortably within the interval $[-x_{\rm max},x_{\rm max}]$. We then compute the energy eigenvalue $E_0$ using a grid spanning the interval with 2048 grid points. The computed $E_0$ must remain unchanged when the number of grid points is decreased to 1024 and increased to 4096. With the converged grid at $m_i$, we repeat the procedure at $m_f$ to ensure that the same grid must guarantee the convergence of $E_0$ at the final mass as well.
}
.
A grid that passed the convergence test at both ends of a stage is used for quantum annealing of that stage.

\subsection{Residual energy}
\label{subsec.residual.QA.linear}

Quantum annealing of a stage is performed by inserting the particular $m_i$ and $m_f$ into the schedule Eq. (\ref{eq.linear schedule definition}), which is then substituted into Eq. (\ref{eq. TDSE }). The resulting Schr\"{o}dinger equation is solved using the WavePacket program package \cite{Schmidt17,Schmidt18,WavePacket}. Atomic units are used, and $m(t)$ is expressed in units of the electron rest mass. Time propagation is performed using the Runge-Kutta algorithm with a time step of $dt=0.1$
\footnote{Convergence of trajectories with respect to $dt$ is checked by halving it (to 0.05) and then seeing that recalculated results remain invariant.}.
The ground state wavefunction at $m_i$ is used as the initial wavefunction at the start of annealing. 

To assess the performance of annealing, we consider the residual energy, defined as 
\begin{equation}
R_{\rm q.a.}(T)=\langle \psi(T) |H(T)|\psi(T) \rangle  - E_0(m_f).
\label{eq.Residual energy QA.definition}
\end{equation}
The first term is the expectation value of the Hamiltonian [given in Eq. (\ref{eq. TDSE })] taken with respect to the wavefunction obtained when annealing finishes at $t=T$. The second term is the ground state energy at the final mass, and is the target energy which annealing seeks to attain. In the limit $T\rightarrow \infty$, the time evolution becomes adiabatic and the residual energy approaches zero. The rate at which $R_{\rm q.a.}(T)$ decays with the total annealing time $T$ is a measure of how quickly the target energy is being attained by the annealing process.

\begin{figure}[h]
\begin{center}
\includegraphics[scale=0.75]{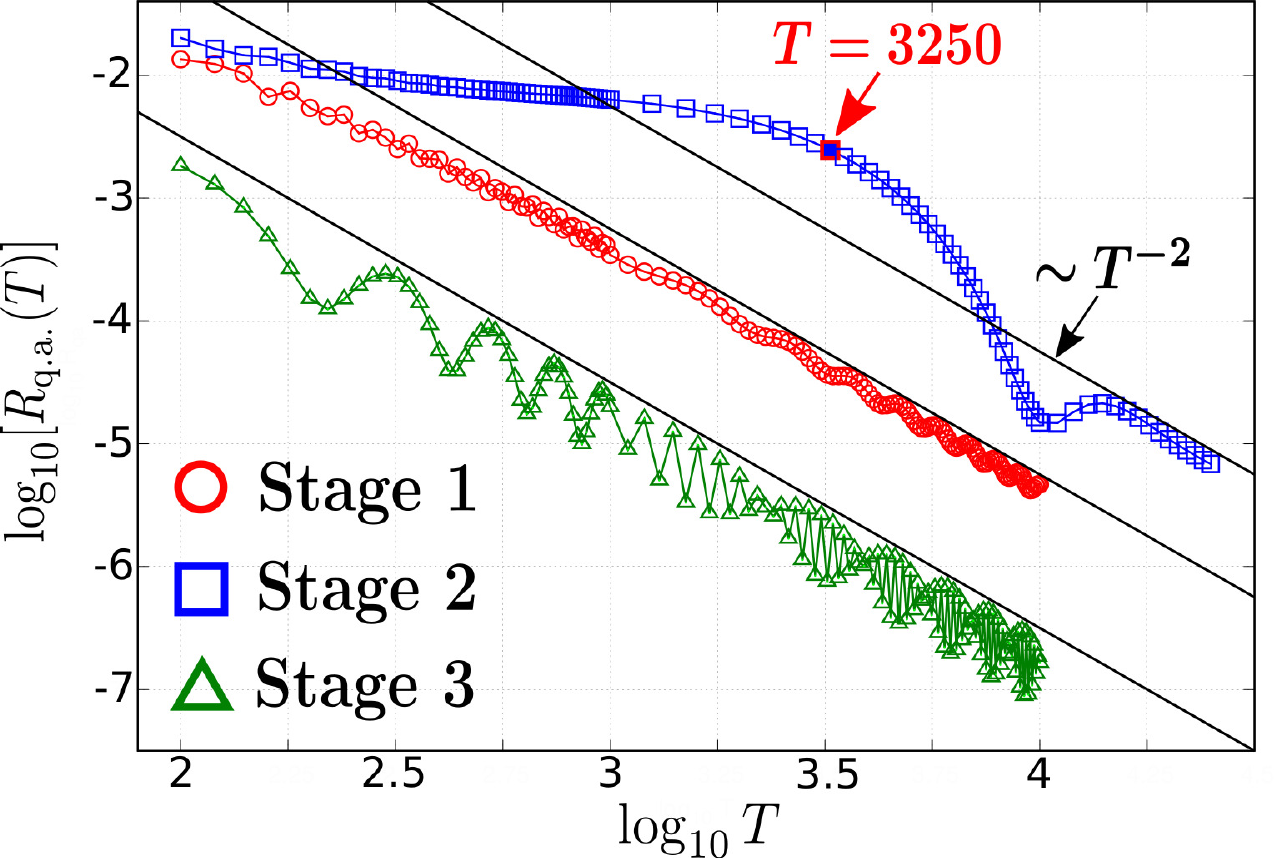}
\caption{Residual energy $R_{\rm q.a.}(T)$ as a function of total annealing time $T$, for the three stages of quantum annealing performed using the  linear schedule Eq. (\ref{eq.linear schedule definition}). The three straight lines (black) are proportional to $T^{-2}$ and show the asymptotic behaviors of the curves. Lines connecting the data points are to guide the eye only.}
\label{fig.Rqa vs T.linear}
\end{center}
\end{figure}

Figure \ref{fig.Rqa vs T.linear} shows $R_{\rm q.a.}(T)$ as a function of $T$ obtained under the linear schedule. The three curves show the results for the three stages discussed earlier. The straight lines (black) are proportional to $T^{-2}$ and indicate the asymptotic behavior of the curves when $T$ is large 
\footnote{Due to the periodic oscillatory behavior of the data points, instead of peforming a least square fit, we have chosen instead to manually adjust the vertical displacement of the line $T^{-2}$ to fit those data points at the crests of the oscillatory curve. We find that this reflects more accurately the observed gradients of the curves.}.
We also performed some additional simulations at other $h_0$-$w_0$ phase points and found that the decay rate $T^{-2}$ holds quite generally in phase space. This decay rate for the linear schedule is in agreement with previous studies on spin systems \cite{Suzuki2005,Morita07}.

\begin{figure}[h]
\begin{center}
\includegraphics[scale=0.7]{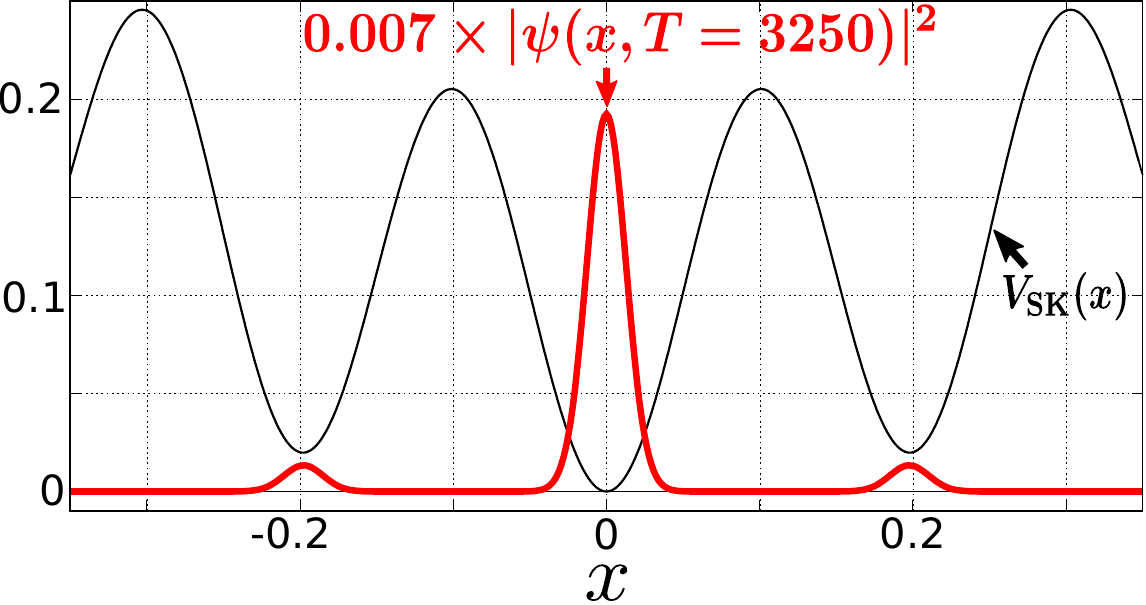}
\caption{An example of a wavefunction that is obtained if annealing is performed too quickly. The red curve shows the final probability density $|\psi(x,T)|^2$ obtained in Stage 2 with $T=3250$, whose residual energy is also indicated in Fig. \ref{fig.Rqa vs T.linear}. The graph of $V_{\rm SK}(x)$ (black) is also shown for comparison. Note the spurious peaks located at the local minima, which are absent in the target ground state [c.f. Fig. \ref{fig. compare quantum classical groundstates . 0.1 and 0.01}(c)].}
\label{fig.badly annealed wavefunction.Stage 2 linear schedule T=3250}
\end{center}
\end{figure}

The curve of Stage 2 (blue squares) is slightly different from the other two stages, first undergoing a transient phase before crossing over into the asymptotic $T^{-2}$ regime. Similar two-phased behavior in the residual energy has also been reported in a two-level system, with the earlier phase being attributed to diabatic transitions and explained by the Landau-Zener theorem \cite{Morita07}. The transient phase here is probably also due to transitions into excited states. Figure \ref{fig.badly annealed wavefunction.Stage 2 linear schedule T=3250} shows an example of the final wavefunction that is obtained in Stage 2, with the corresponding residual energy indicated in Fig. \ref{fig.Rqa vs T.linear} (arrowed $T=3250$). For comparison, the target ground state is shown in Fig. \ref{fig. compare quantum classical groundstates . 0.1 and 0.01}(c). One sees that the annealed wavefunction exhibits two spurious peaks at the neighboring local minima, which are absent in the target wavefunction. Furthermore, we found that the first excited level in the `flat gap' region (c.f. Fig. \ref{fig. gap vs m }) is doubly-degenerate, spanned by two compact gaussians each centered at one of the two local minima. Hence, the spurious peaks in the annealed wavefunction in Fig. \ref{fig.badly annealed wavefunction.Stage 2 linear schedule T=3250} are most likely caused by diabatic transition into the first excited level when the annealing traverses the `flat gap' region.

It is interesting to see that in quantum annealing, parts of the wavefunction can become trapped near local minima if the annealing time is not long enough. Normally, this may be what one would expect more in simulated annealing. This similarity between quantum and simulated annealing is inspiring from a physics point of view, especially since two annealing protocols obey very different rules. The mechansim underlying the trapping in simulated annealing is fairly well-understood. On the other hand, we might try to explain the trapping observed for quantum annealing using Landau-Zener transition. In the case of $V_{\rm SK}(x)$, we have seen that the system encounters a region of `flat gap' during the course of annealing. This may pose a problem for the Landau-Zener theory as there is no well-defined avoided crossing, which is one of the basic tenets of the Landau-Zener theory. Hence, although the physical process underlying the trapping observed here might be a diabatic transition, the detailed mechanism could differ from the traditional Landau-Zener one. 

\subsection{Time evolution of the wavefunction and the approach to adiabaticity}
\label{subsec.time evolution.QA.linear}

\begin{figure}[h]
\begin{center}
\includegraphics[scale=0.8]{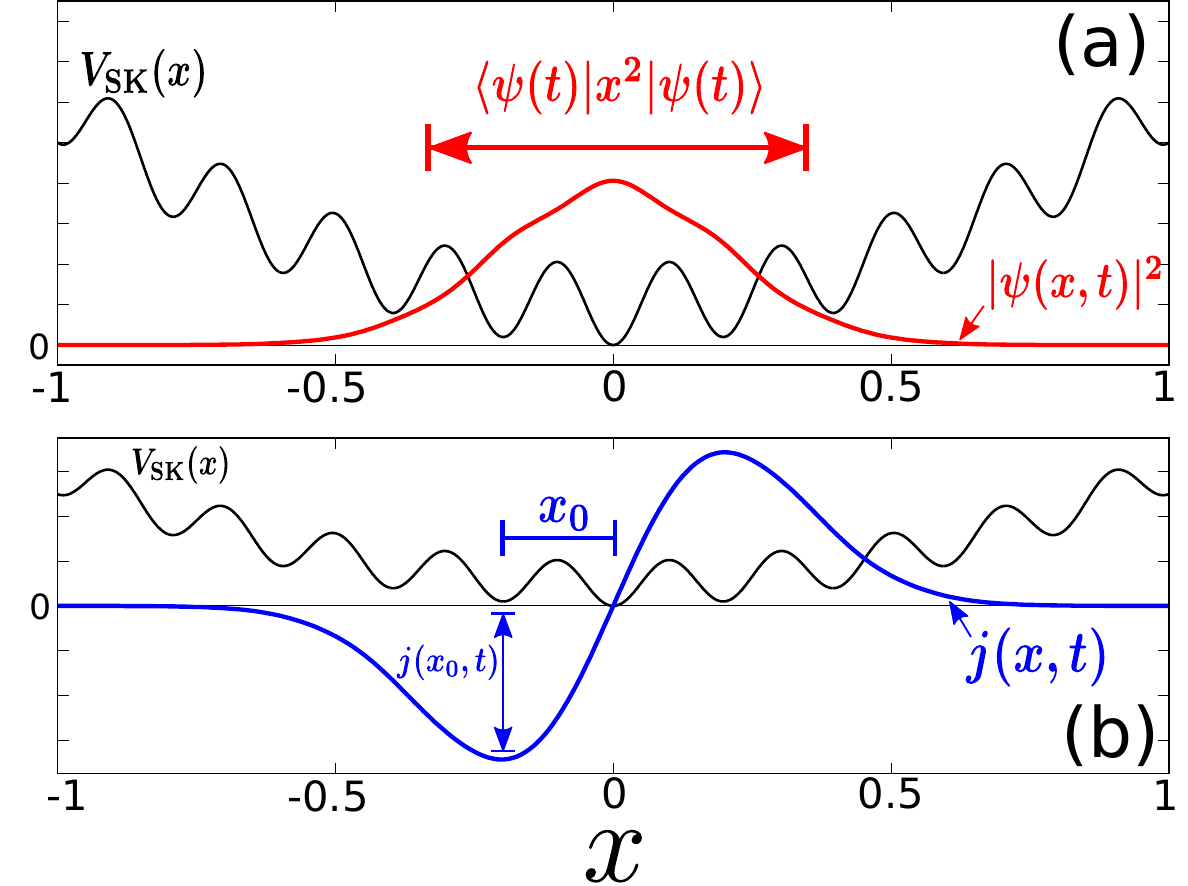}
\caption{Snapshot of the wavefunction at time $t=44$ during annealing in Stage 1 with $T=560$.  (a) The probability density $|\psi(x,t)|^2$, with width given by the expectation value of $x^2$. (b) The probability current $j(x,t)$, with amplitude $j(x_0,t)$ at $x_0$. In both panels, the graph of $V_{\rm SK}(x)$ has been rescaled vertically for presentation purposes.}
\label{fig.explaining width and amplitude}
\end{center}
\end{figure}

\begin{figure}[h]
\begin{center}
\includegraphics[scale=0.7]{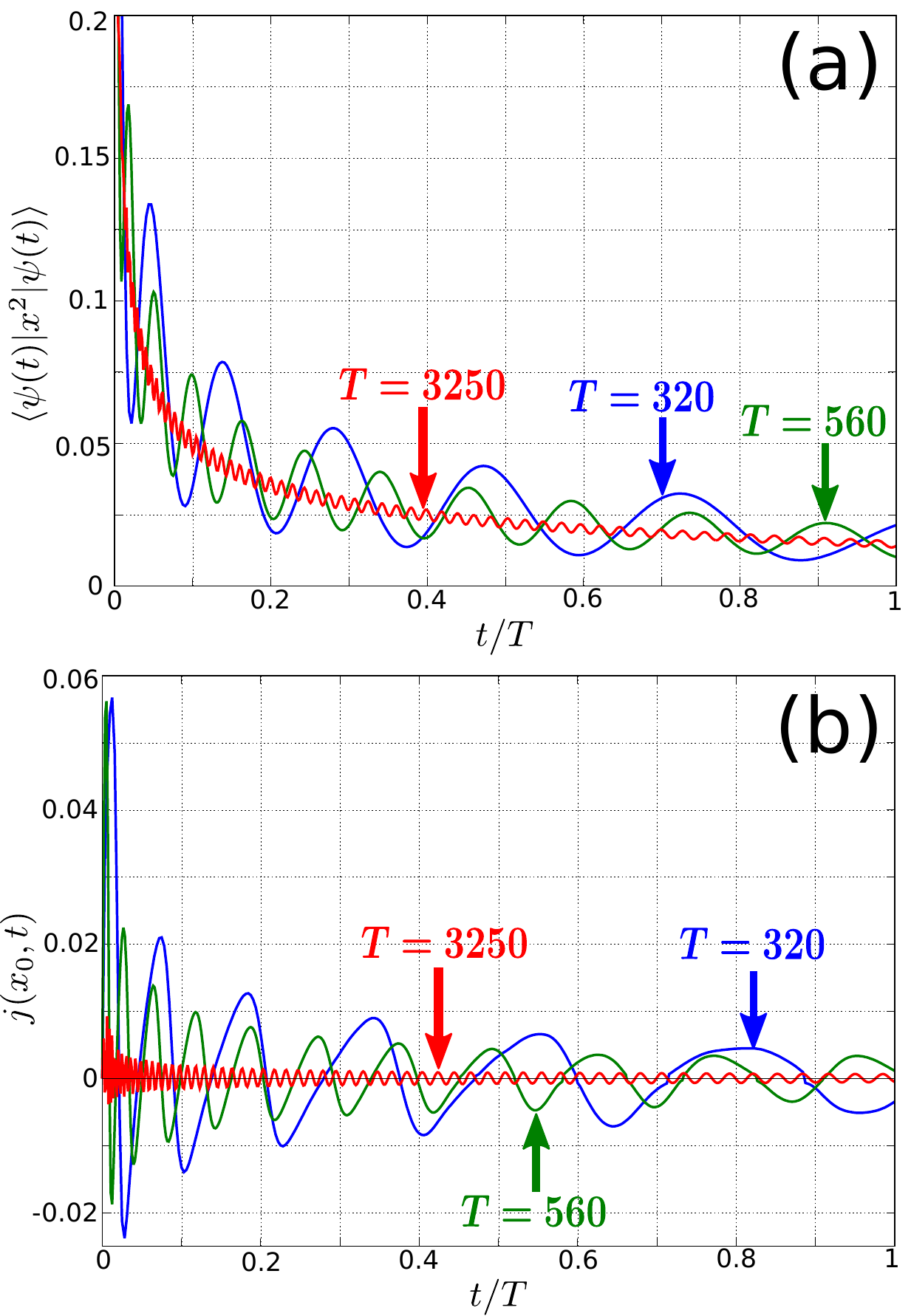}
\caption{Time evolutions of (a) the width $\langle \psi(t)|x^2|\psi(t)\rangle$ and (b) the current amplitude $j(x_0,t)$ of the wavefunction in Stage 1 under the linear schedule. The graphs of $T=320, 560$, and 3250 are compared to show the effects of increasing annealing time. In panel (b), the diminishing of $j(x_0,t)$ towards zero with increasing $T$ indicates the approach to adiabaticity.}
\label{fig.time evoluton.width and amplitude.stage1.linear.various Ts}
\end{center}
\end{figure}

We now discuss the time evolution of the wavefunction during the course of annealing. Figure \ref{fig.explaining width and amplitude}(a) shows a snapshot of the probability density at time $t=44$ in Stage 1, with $T=560$. The single-modal form is quite representative of the shape of the wavefunction in Stage 1, so we can express its temporal behavior more succinctly using the width $\langle \psi(t)|x^2|\psi(t)\rangle$, which is illustrated in Fig. \ref{fig.explaining width and amplitude}(a). Figure \ref{fig.time evoluton.width and amplitude.stage1.linear.various Ts}(a) shows the time evolution of the width in Stage 1 for several values of $T$. Overall, one sees that all three graphs oscillate, with the amplitude of oscillation decaying with the passage of time. Among the graphs themselves, as $T$ increases, the oscillation becomes suppressed to yield a smoother decay.

Although informative, looking at the width alone gives us rather limited understanding of the behavior of the wavefunction. More insight can be gained by looking at the so-called probability current, defined as 
\begin{equation}
j(x,t)=\frac{\hbar}{2 m(t) i}\left( \psi^* \frac{\partial \psi}{\partial x}  - \psi \frac{\partial \psi^*}{\partial x} \right),
\label{eq.j(x,t).definition}
\end{equation}
where $\psi^*(x,t)$ is the complex conjugate of $\psi(x,t)$, $i=\sqrt{-1}$, and $m(t)$ is again the time-dependent mass. The current $j(x,t)$ is a real-valued quantity, and indicates the direction in which the probability is flowing at a point in space 
\footnote{For the plane wave $e^{ikx}$, one has $j(x,t)=\frac{\hbar k}{m}=\frac{p}{m}=v$, which is simply the velocity of the particle. Hence, the probability current can be interpreted as a generalization of the velocity field, indicating both the speed and direction of motion of the particle.
}
.
Figure \ref{fig.explaining width and amplitude}(b) shows the probability current corresponding to the probability density shown in panel (a). It is seen that the current is positive (negative) when $x>0$ ($x<0$), meaning that probability is flowing outwards away from the global minimum at the origin. With respect to the purpose of optimization, such an instantaneous behavior of the wavefunction should be considered unfavorable because the particle is diffusing away from the optimal solution at $x=0$.

As with the width of the probability density, let us also introduce a parameter associated with $j(x,t)$ to monitor the time evolution of the current. Define the amplitude as the largest displacement of $j(x,t)$ from zero in the region $x<0$. This is illustrated in Fig. \ref{fig.explaining width and amplitude}(b), where the amplitude is $j(x_0,t)$ at the point $x_0$. The amplitude can take on negative values, with the sign indicating the direction of probability flow.

Figure \ref{fig.time evoluton.width and amplitude.stage1.linear.various Ts}(b) shows the time evolutions of the current amplitude $j(x_0,t)$ associated with the time evolutions of the width shown in panel (a). We see that the amplitude oscillates in time between positive and negative values, meaning that the probability current flow towards and away from the global minimum in a periodic manner. This is consistent with the pulsating behavior exhibited by the width. Another feature is that as $T$ increases, the amplitude tends to zero, meaning that the current itself approaches zero both spatially, as well as temporally throughout the entire duration of annealing. As $j(x,t)$ is identically zero for stationary states, this is a sign of the annealing process approaching adiabaticity. Viewed differently, even for the moderate value of $T=560$, where the residual energy approaches the asymptotic $T^{-2}$ law for Stage 1 as seen in Fig.~\ref{fig.Rqa vs T.linear} (at $\log_{10}T=2.748$), the current keeps oscillating, see Fig.~\ref{fig.time evoluton.width and amplitude.stage1.linear.various Ts}(b), implying a diabatic evolution.

So far, we have focused our discussion on Stage 1. The probability current in Stage 2 is slightly more complicated. An example is shown in Fig. \ref{fig.snapshot of j(x,t).stage2.linear.t=10.T=320}, where a snapshot of $j(x,t)$ at $t=10$ with $T=320$ is shown.  One sees that the current direction varies spatially, unlike the spatially-coherent flow we have seen in Fig. \ref{fig.explaining width and amplitude}(b). Despite such differences, the overall behaviors in Stages 2 and 3 are generally quite similar to Stage 1. For completeness, these results are discussed in Appendix \ref{app.time evolution of current in stages 2 and 3}.


\section{Quantum annealing with nonlinear schedules}
\label{sec. nonlinear schedule}

\subsection{Nonlinear annealing schedules}

In the previous section, we have seen that asymptotically the residual energy decays with the total annealing time as $T^{-2}$. Analysis related to the adiabatic theorem shows that this scaling is largely a result of annealing with the linear schedule Eq. (\ref{eq.linear schedule definition}) \cite{Suzuki2005,Morita07}. Nonlinear schedules have been proposed in which the residual energy decays much faster with the annealing time, leading to better efficiency \cite{Morita07}.
In this section, we apply them to the potential $V_{\rm SK}(x)$.

Nonlinear schedules can be obtained by generalizing Eq. (\ref{eq.linear schedule definition}) as
\begin{equation}
m^{(n)}(t)=\left( m_f- m_i \right)f_n(t/T) + m_i,
\label{eq.nonlinear schedules. definition}
\end{equation}
where $n$ denotes the order of the schedule, and the linear function $t/T$ is replaced by higher-order functions $f_n(t/T)$. Details concerning the derivation of $f_n$ can be found in Ref. \cite{Morita07}. Physically speaking, these functions are designed such that annealing is constrained to proceed slowly both at the beginning as well as at the end of the process. Earlier studies on spin systems revealed that the residual energy decays with annealing time as $T^{-2n}$, so higher-order schedules generally lead to better annealing performances. In the following, we shall investigate the performances of the functions $f_1(s)$ to $f_4(s)$ given in Ref. \cite{Morita07},
\begin{align}
f_1(s)&=s, ~f_2(s)=s^2(3-2s), \\
f_3(s)&=s^3(10-15s+6s^2), ~f_4(s)=s^4(35-84s+70s^2-20s^3).
\label{eq.nonlinear4}
\end{align}

\subsection{Residual energy: faster decay under nonlinear schedules}

\begin{figure}[h]
\begin{center}
\includegraphics[scale=0.5]{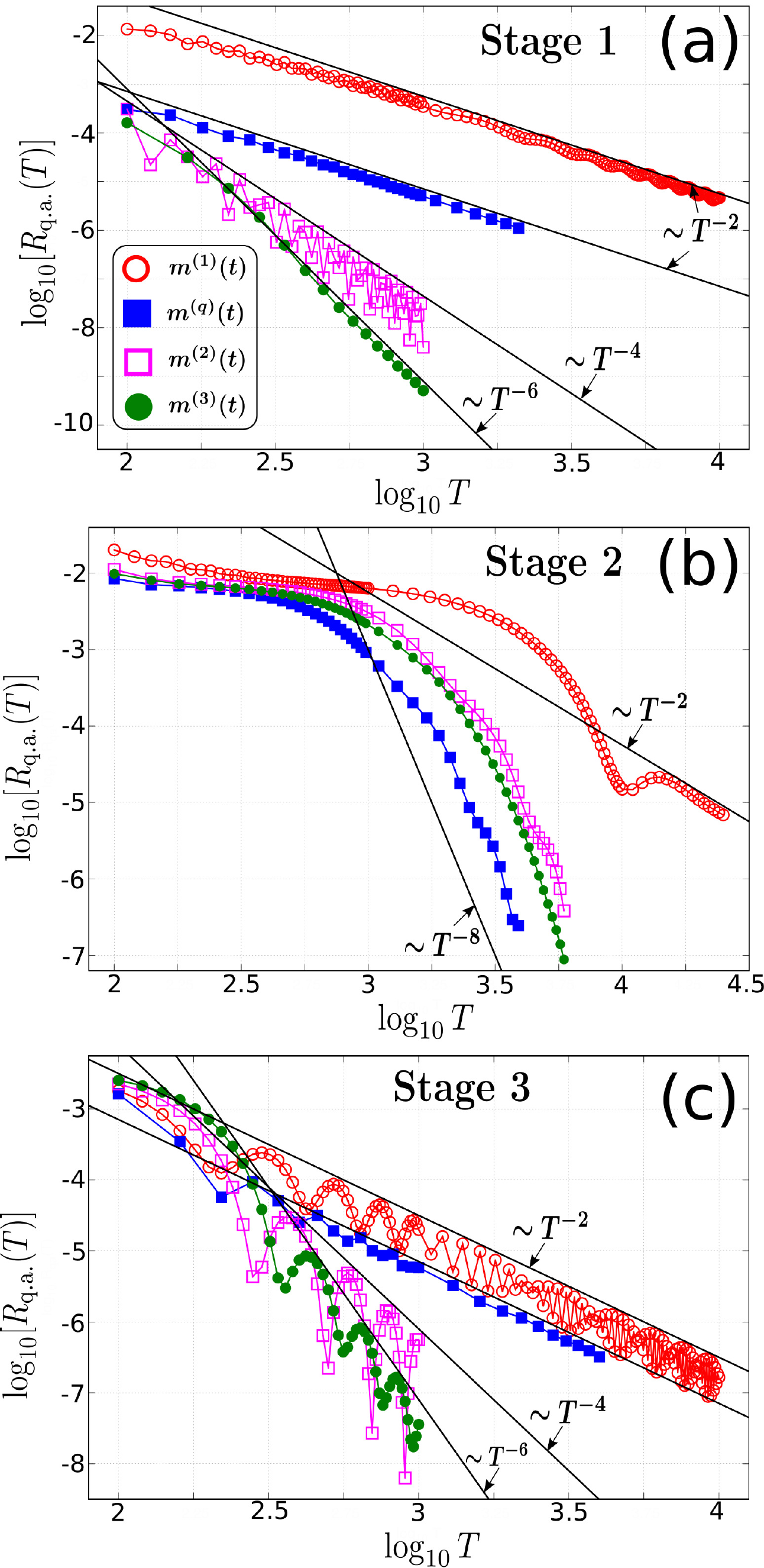}
\caption{Comparing the performances of linear and nonlinear schedules in quantum annealing. Panels (a), (b), and (c) show the $R_{\rm q.a.}(T)$ versus $T$ curves in Stage 1, 2, and 3, respectively. Legend for the schedules shown in panel (a) also applies to (b) and (c). The straight lines (black) are proportional to $T^{-2n}$ ($n=1$ to 4) and indicate the asymptotic behaviors of the curves. Lines connecting the data points are to guide the eye only.}
\label{fig.Rqa vs T.comparing linear and nonlinear schedules}
\end{center}
\end{figure}

Figure \ref{fig.Rqa vs T.comparing linear and nonlinear schedules} compares the residual energies obtained under the schedules given by Eq. (\ref{eq.nonlinear schedules. definition}), with $n=1$ to 3. The results from Stages 1 to 3 are shown in panels (a) to (c) of Fig. \ref{fig.Rqa vs T.comparing linear and nonlinear schedules}, respectively. The general behaviors in Stages 1 and 3 are qualitatively quite similar. It is seen that for an $n$th-order schedule the residual energy decays as $T^{-2n}$, verifying the theoretical prediction of Ref.~\cite{Morita07}. To emphasize the importance of varying the schedule slowly both at the start as well as towards the end of annealing, we also considered a straightforward quadratic schedule $m^{(q)}(t)$, partly motivated by Ref.~\cite{Stella05}, which is obtained by simply substituting $f_n$ with $(t/T)^2$ in Eq. (\ref{eq.nonlinear schedules. definition}). This schedule proceeds slowly only at the start but not at the end of annealing. In panels (a) and (c), we see that the residual energy of $m^{(q)}(t)$ (solid blue squares) decays only as $T^{-2}$. That both linear and quadratic schedules should show the same asymptotic decay rate can also be explained based on the theoretical analysis of Ref. \cite{Morita07}.

The situation in Stage 2 is different from the other two in that the residual energy decays in the manner of the transient phase discussed in Sec. \ref{subsec.residual.QA.linear}. Note that this is in the domain of strongly diabatic transitions and so is not the regime in which the function $f_n$ were originally designed for. Nevertheless, it is seen from panel (b) of Fig.~\ref{fig.Rqa vs T.comparing linear and nonlinear schedules} that the residual energies of $m^{(2)}(t)$ and $m^{(3)}(t)$ decrease faster than that of $m^{(1)}(t)$. However, we also found that the curve for $m^{(4)}(t)$ (not shown) is almost exactly the same as that of $m^{(3)}(t)$, so we should be cautious about generalizing the improvement in performance seen for $n=2$ and 3 to higher orders. We were unable to pursue the curves of $m^{(2)}(t)$ and $m^{(3)}(t)$ to larger $T$ to examine the asymptotic crossover because we have exhausted the numerical precision offered by the simulation package. Within the time span we could study as shown in  Fig.~\ref{fig.Rqa vs T.comparing linear and nonlinear schedules}(b), diabatic processes lead to significantly faster convergence than $T^{-2}$.  Interestingly, the quadratic schedule gives the best performance here, unlike in the other two stages. Overall, nonlinear schedules generally show improvement over the linear one in this stage. Interpretation of the results, however, is not as clear-cut as in Stages 1 and 3.

\subsection{Persistent tunneling current during quantum annealing}
\label{subsec.persistent tunneling under nonlinear schedules}

Earlier in Sec. \ref{subsec.time evolution.QA.linear}, we have seen that under the linear schedule, increasing the annealing time suppresses of the probability current and leads to a more adiabatic process. Conventionally, adiabaticity is considered to be closely related with how the global minimum is ultimately to be attained in quantum annealing. With nonlinear schedules, however, we are now allowed to explore the schedule's order $n$ as a new parameter dimension while keeping the annealing time constant. We shall see that this leads to tunneling, as opposed to adiabaticity, as a dominant mechanism for quantum annealing to arrive at the optimal solution.

Let us focus our discussion on Stage 1. The probability currents under nonlinear schedules also have the spatially-coherent form of Fig. \ref{fig.explaining width and amplitude}(b) (no node in each of the region $x>0$ and $x<0$), so once again we can use the amplitude $j(x_0,t)$ to monitor their temporal behaviors. Figure \ref{fig.compare  j(x0,t) various schedules.stage1.T560} shows the time evolutions of $j(x_0,t)$ obtained under the various schedules, with annealing time $T=560$. The salient feature that is common to all the nonlinear schedules and which distinguishes them from the linear one is that the amplitude is consistently positive throughout the entire duration of annealing. The probability current is therefore always flowing towards the origin, and no current ever flows away. This is a highly diabatic process in which the optimization is targeted at the global minimum. In stark contrast, the current of the linear schedule oscillates in an unguided manner, resulting in inefficient annealing.

\begin{figure}[h]
\begin{center}
\includegraphics[scale=0.7]{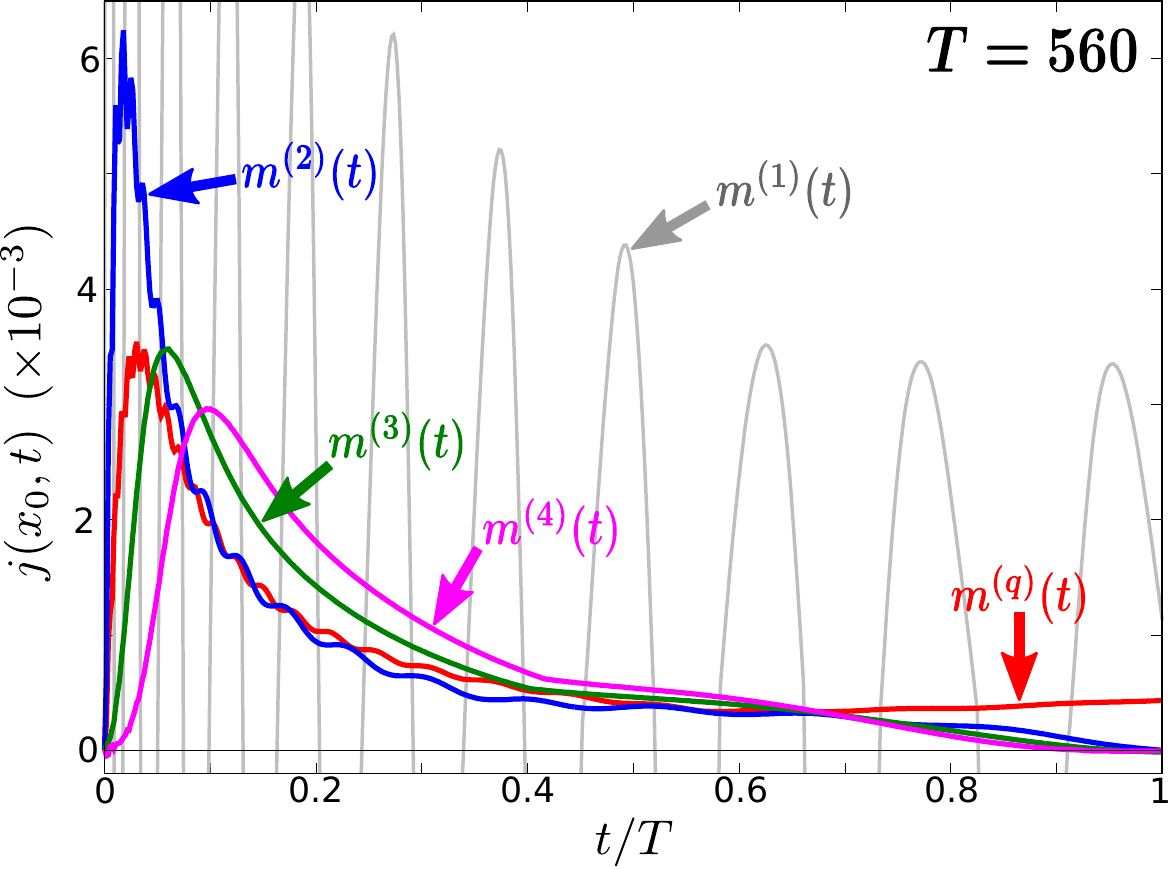}
\caption{Comparing the time evolutions of $j(x_0,t)$ under different annealing schedules $m^{(n)}(t)$ in Stage 1. The oscillatory behavior under the linear schedule $m^{(1)}(t)$ (gray) has also been shown in Fig. \ref{fig.time evoluton.width and amplitude.stage1.linear.various Ts}(b) (green). For nonlinear schedules, the $j(x_0,t)$ are strictly positive, indicating a persistent current directed towards the global minimum throughout the entire annealing.}
\label{fig.compare  j(x0,t) various schedules.stage1.T560}
\end{center}
\end{figure}

\begin{figure}[h]
\begin{center}
\includegraphics[scale=0.75]{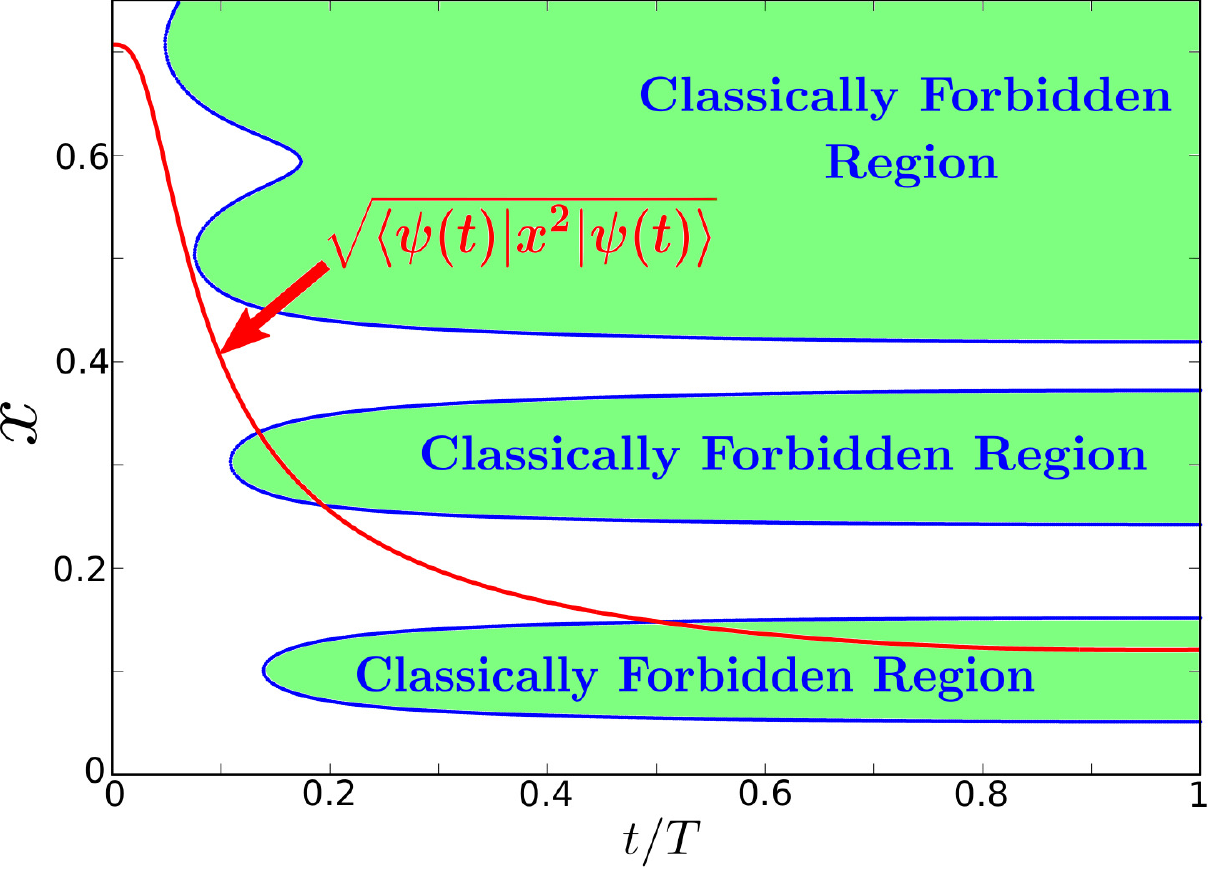}
\caption{Tunneling exhibited by the wavefunction when annealing under a nonlinear schedule. Tics along the vertical axis represent the domain $x>0$ and also indicate the time evolution of $\sqrt{\langle \psi(t)|x^2|\psi(t) \rangle}$ in Stage 1 under $m^{(3)}(t)$ with $T=560$ (red curve). The (green) shaded areas indicate the classically forbidden regions where $\langle\psi(t)|H(t)|\psi(t)\rangle<V_{\rm SK}(x)$. The shaded areas lying below the red curve are where the wavefunction exhibits tunneling.}
\label{fig.explain tunneling using standard deviation.}
\end{center}
\end{figure}

Apart from being persistent, the current of nonlinear schedules also has the additional feature of exhibiting tunneling. To explain, let us first note from Fig. \ref{fig.explaining width and amplitude} that the spatial extent of both the probability density and the current in the domain $x>0$ can be approximated by the width $\sqrt{\langle \psi(t)|x^2|\psi(t)\rangle}$. The red curve in Fig. \ref{fig.explain tunneling using standard deviation.} shows the time evolution of $\sqrt{\langle \psi(t)|x^2|\psi(t)\rangle}$ for the schedule $m^{(3)}(t)$ in Stage 1 with $T=560$. The (green) shaded areas show the classically forbidden regions where the total energy $\langle \psi(t)|H(t)|\psi(t) \rangle$ is less than the potential energy $V_{\rm SK}(x)$. Shaded areas lying below the red curve are regions where the current resides in classically forbidden regimes, and these currents are tunneling in nature. Hence, nonlinear schedules induce persistent currents that tunnel across potential barriers to greatly improve the efficiency of annealing. This is in contrast to the linear schedule which, on the contrary, seeks to suppress the flow of current in order to bring about a more adiabatic annealing process.

So far, our discussion has focused on Stage 1. For Stage 2, we mentioned earlier that under the linear schedule the current is not spatially-coherent (c.f. Fig. \ref{fig.snapshot of j(x,t).stage2.linear.t=10.T=320}). Under nonlinear schedules, however, the current resumes a spatially-coherent form and so we can again use the amplitude $j(x_0,t)$ as a monitoring parameter to study its temporal evolution. Despite the different behavior of their residual energies, it was found that the observations reported for Stage 1 are valid for Stage 2 as well. For completeness, these results are summarized in Appendix \ref{app.time evolution of current in stages 2 and 3}.

The situation in Stage 3 is quite similar to that of Stage 1, with the current showing persistence under the nonlinear schedules. Tunneling, however, is not a prominent feature in Stage 3 as the wavefunction is already well-localized within the global minimum [c.f. Fig. \ref{fig. compare quantum classical groundstates . 0.1 and 0.01}(c)]. The optimization is therefore taking place within an effective quadratic potential, which is a relatively simple problem for quantum annealing.

\section{Simulated annealing using a linear schedule}
\label{sec.SA.linear schedules}
We next study simulated annealing, first under a linear annealing schedule for comparison with the quantum counterpart.
\subsection{Annealing protocol}
We perform simulated annealing numerically using the Metropolis algorithm. Let the position of the particle be $x_t$ at time $t$. A new position $x'$ is proposed as 
\begin{equation}
x'=x_t+\Delta x
\label{eq.proposed move}
\end{equation}
where $\Delta x$ is a random number drawn from the interval $[-\frac{s}{2},\frac{s}{2}]$ of width $s$. The quantity $s$ is called the Monte Carlo step size and it determines how far the particle is allowed to jump with each move. The proposed position $x'$ is then accepted with probability
\begin{equation}
P_{\rm accept}=\min \left\{ 1 , e^{-\beta(t)\left[ V_{\rm SK}(x') - V_{\rm SK}(x_t) \right]} \right\},
\label{eq.metropolis acceptance}
\end{equation}
where the temperature is provided by the annealing schedule $\beta(t)$, which will be specified later. After $x_t$ is updated, the time $t$ advances by a time step $dt$ and the Metropolis algorithm is repeated 
\footnote{The simulated annealing calculations in Secs. \ref{sec.SA.log schedule} and \ref{sec.SA.linear schedules} are performed with $dt=0.1$, to be consistent with the time step used in quantum annealing.
}. 

To compute averages, we work with an ensemble of $N$ particles where each of them evolves independently in time according to the Metropolis algorithm. The average $\langle V_{\rm SK}(x)\rangle_t$ at time $t$ is calculated via the ensemble average as
\begin{equation}
\langle V_{\rm SK}(x)\rangle_t = \lim_{N\rightarrow \infty}  \frac{1}{N}\sum_{\nu=1}^{N} V_{\rm SK}(x_t^{\nu}),
\label{eq.ensemble average of Vsk}
\end{equation}
where $x^{\nu}_t$ is the position of the $\nu$th particle at time $t$. In practice, one works with a large but finite $N$, typically $N=10^7$ and $10^9$. The ensemble average then becomes an approximation, subjected to errors due to finite-size fluctuations. 
\subsection{Linear schedule and stages of simulated annealing}
Let us consider the following linear form for $\beta(t)$
\begin{equation}
\beta^{(1)}(t)=\left(\beta_f-\beta_i \right)  \left(\frac{t}{T}\right) + \beta_i.
\label{eq.sa.linear schedule.definition}
\end{equation}
As in quantum annealing, we restrict ourselves to the time interval $0\le t \le T$, where $T$ again denotes the total annealing time. The parameters $\beta_i$ and $\beta_f$ are the inverse temperatures at the initial ($t=0$) and final ($t=T$) times, respectively. The schedule $\beta^{(1)}(t)$ is the simulated annealing counterpart to the mass schedule $m^{(1)}(t)$ that we have used in quantum annealing.

We would like to perform the simulated annealing calculations in such a way that they can be compared to the results of quantum annealing obtained in Sec. \ref{sec. linear schedule}. To enable a fair comparison, we use energy as the criterion. The boundary parameters $\beta_i$ and $\beta_f$ in Eq. (\ref{eq.sa.linear schedule.definition}) are chosen in such a way that the initial and final energies of a stage in simulated annealing are approximately the same as the initial and final energies of the corresponding stage in quantum annealing. These boundary parameters $\beta_l$ and their energies $U(\beta_l)$ are summarized in Table \ref{tab.summary of boundaries of annealing stages}, alongside their corresponding quantum counterparts $m_l$ and $E_0(m_l)$. To facilitate our subsequent discussions, let us call annealing from $\beta_a$ to $\beta_b$ Stage A, that from $\beta_b$ to $\beta_c$ Stage B, and that from $\beta_c$ to $\beta_d$ Stage C. The equilibrium state properties at the stage boundaries have been discussed in Sec. \ref{sec. Compare statics} [c.f. Figs. \ref{fig. compare quantum classical groundstates . 0.6}(b), \ref{fig. compare quantum classical groundstates . 0.1 and 0.01}(b), and \ref{fig. compare quantum classical groundstates . 0.1 and 0.01}(d)]. A graphical perspective of the three stages is also given in Fig. \ref{fig. E0 and U vs m and beta}(b), where they are superposed upon the curve of the internal energy $U(\beta)$. It is seen that the stages are positioned in the non-trivial regime of the system exhibiting `excess energy'.

\subsection{Residual energy}

The annealing performance is assessed by considering the residual energy 
\begin{equation}
R_{\rm s.a.}(T)=\langle V_{\rm SK}(x) \rangle_{t=T}-\langle V_{\rm SK}(x) \rangle_{\beta_f}~.
\label{eq.residual energy.sa.definition}
\end{equation}
The first term is the average potential energy obtained when annealing ends at $t=T$. The second term is the average potential energy at equilibrium at the final temperature [c.f. Eq. (\ref{eq.average potential energy.definition})], and is the target energy that annealing seeks to attain. The quantity $R_{\rm s.a.}(T)$ is the simulated annealing counterpart to the residual energy $R_{\rm q.a.}(T)$ considered in quantum annealing.

\begin{figure}[h]
\begin{center}
\includegraphics[scale=0.75]{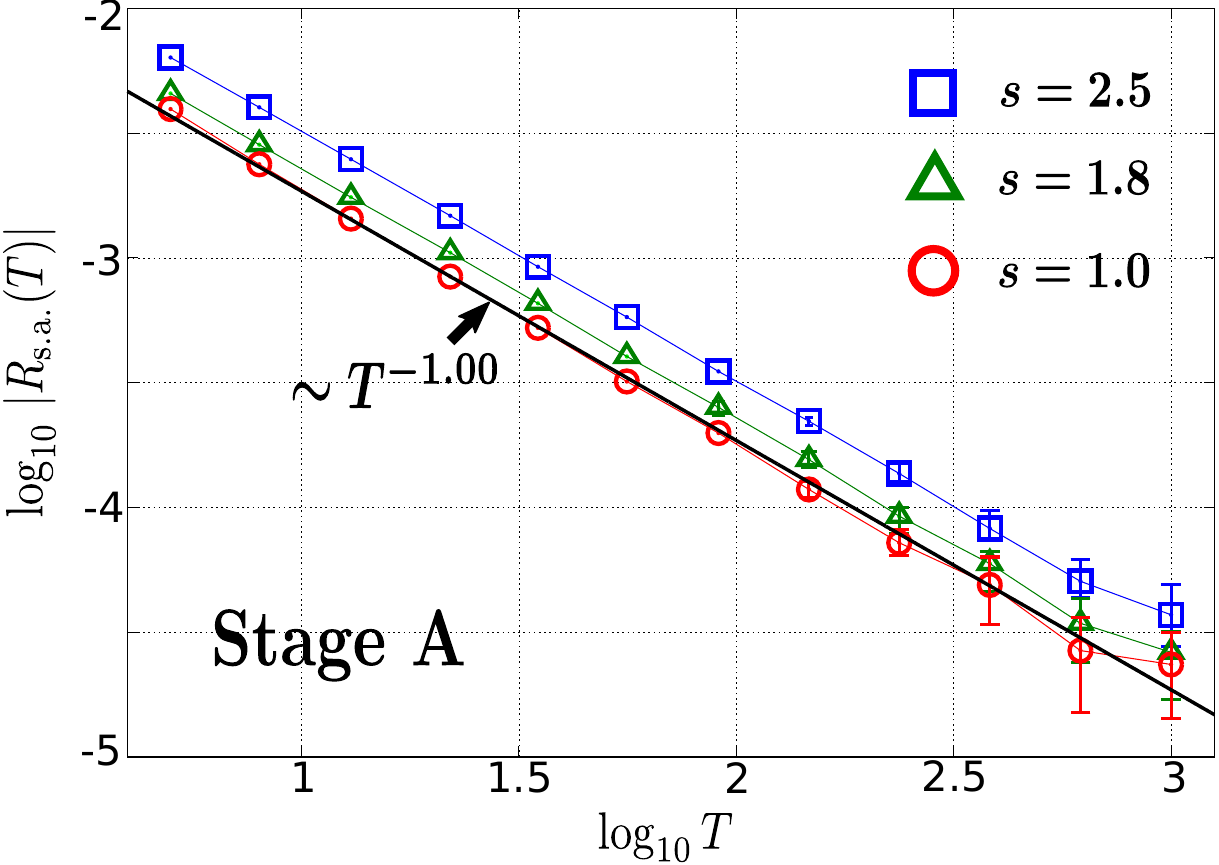}
\caption{Decrease of residual energy $R_{\rm s.a.}(T)$ with total annealing time $T$ for simulated annealing, performed under the linear schedule Eq. (\ref{eq.sa.linear schedule.definition}) in Stage A. Results for three step sizes $s$ are shown. All data points are obtained by averaging over an ensemble of $N=10^9$, and the error bars indicate finite-$N$ fluctuations
\footnote{The error bars are generated as follows. The entire ensemble of size $10^9$ is divided, arbitrarily, into ten sub-ensembles each of size $10^8$. The residual energy of each sub-ensemble is calculated, and the largest and smallest ones obtained are indicated by the upper and lower bounds of the error bar.
}.
The solid line (black) is obtained by fitting a straight line to the data points of $s=1$. Thin lines connecting data points are to guide the eye only.}
\label{fig.sa. residual energy versus T. Stage A}
\end{center}
\end{figure}

Figure \ref{fig.sa. residual energy versus T. Stage A} shows the behavior of $R_{\rm s.a.}(T)$ as a function of $T$ in Stage A under the linear schedule. We experimented with a wide range of step sizes and some examples of the residual energy curves we obtained are shown in the figure. There exists an optimum step size that gives the lowest-lying $R_{\rm s.a.}(T)$ curve, and in Stage A this step size was found to be $s=1$ (red circles). The solid line (black) is obtained by fitting a straight line to the data points of $s=1$. We see that the residual energy decreases as $T^{-1}$. This scaling of $R_{\rm s.a.}(T)$ with $T$ is quite general. We identified the optimum step sizes in Stages B and C as well, and the lowest-lying $R_{\rm s.a.}(T)$ curve in all three stages are summarized in Fig. \ref{fig.sa. residual energy vs T. Stages A,B,C. also trapping results}, shown by the three slanting curves. The two solid lines (black) are obtained by fitting a straight line to the data points of Stages B and C. It is seen that the residual energy also decreases as $T^{-1}$ in these two stages.

\begin{figure}[h]
\begin{center}
\includegraphics[scale=0.75]{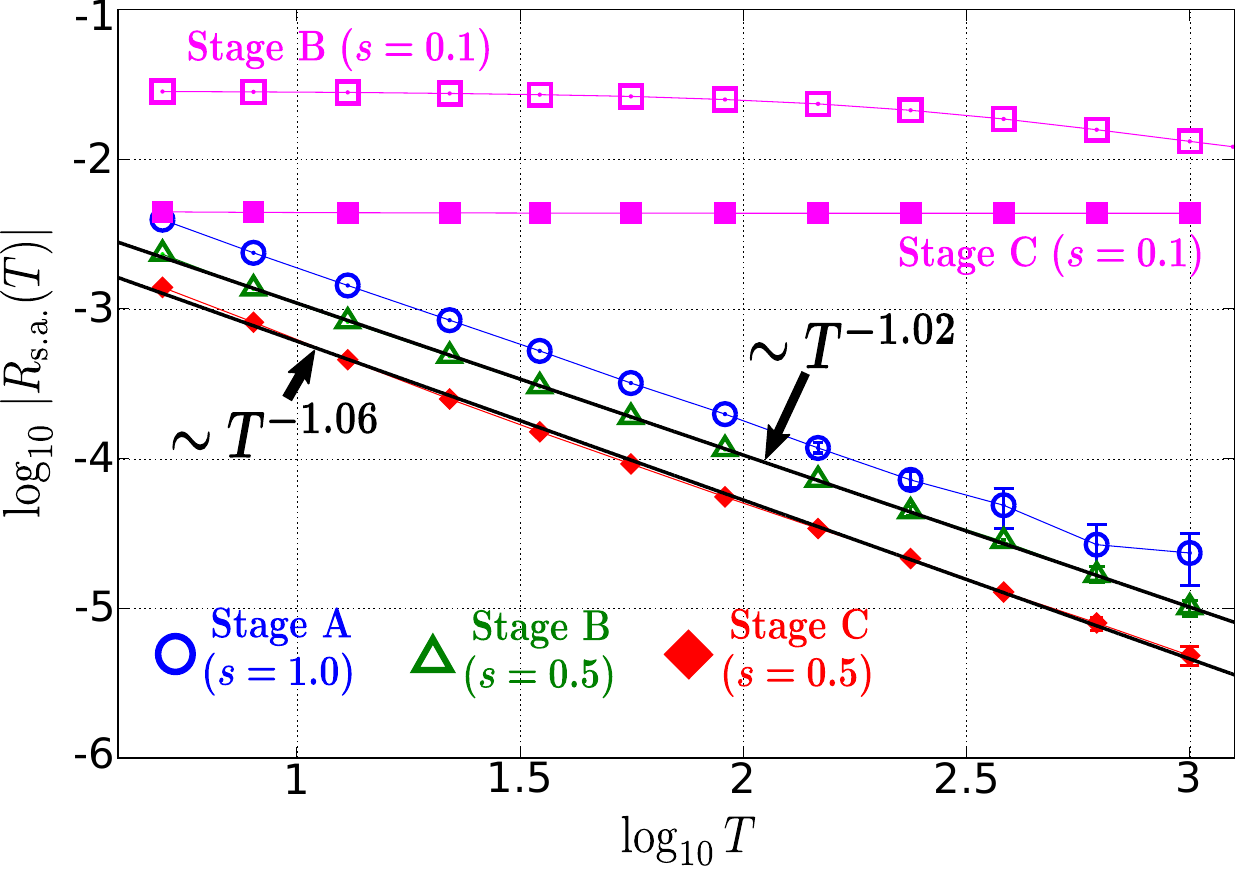}
\caption{Residual energy $R_{\rm s.a.}(T)$ versus annealing time $T$ curves for simulated annealing under the linear schedule. The three slanting curves (circle, triangle, diamond) show the results for Stages A to C, with each curve obtained using the optimum step size of that stage.  The solid straight lines (black) are obtained by fitting to the data points of Stages B and C. The two curves at the top (magenta squares) exhibit slow decay due to trappings by local minima, which happens when the step size is small ($\frac{s}{2}<w_0$). All data points are obtained by averaging over an ensemble of $N=10^9$, and the error bars are generated in the same way as in Fig. \ref{fig.sa. residual energy versus T. Stage A}. Thin lines connecting data points are to guide the eye only.}
\label{fig.sa. residual energy vs T. Stages A,B,C. also trapping results}
\end{center}
\end{figure}

The scaling of $T^{-1}$ is fairly robust in the sense that it is attainable over quite a wide range of step sizes. However, in Stages B and C, we also found that if the step size falls below the inter-minima distance of $V_{\rm SK}(x)$ (i.e. $\frac{s}{2}<w_0$), then the $T^{-1}$ scaling will no longer be achieved. In Fig. \ref{fig.sa. residual energy vs T. Stages A,B,C. also trapping results}, the two uppermost curves (magenta squares) show the residual energies obtained in these two stages with the annealing performed using $s=0.1$. It is seen that they decrease much slower than the previous curves. The reason for the slow decay is trapping by local minima, and can be understood by looking at the initial Boltzmann distributions of these two stages, shown in Figs. \ref{fig. compare quantum classical groundstates . 0.1 and 0.01}(b) and \ref{fig. compare quantum classical groundstates . 0.1 and 0.01}(d). The probability channels connecting the global minimum to the two adjacent local ones are suppressed, so if the step size is smaller than the inter-minima distance $w_0$, particles at the local minima will need a long time to hill-climb the potential barrier in order to reach the global minimum, possibly leading to a logarithmic decrease of the residual energy as suggested by Shinomoto and Kabashima \cite{Shinomoto91}. See also discussions in the next section. Viewed differently, larger step sizes satisfying $\frac{s}{2}>w_0$ may be regarded to realize quasi-global searches across energy barriers without hill-climbing processes, leading to a faster polynomial decrease of the residual energy. It is noteworthy that, even with such a quasi-global search, the decrease rate $T^{-1}$ is slower than the quantum counterpart $T^{-2}$.

One might inquire if it is possible to obtain a better scaling than $T^{-1}$ by making some adjustments to the annealing algorithm. We tried a simple quadratic schedule where the linear term $t/T$ in Eq. (\ref{eq.sa.linear schedule.definition}) is squared. We also experimented with a variant algorithm in which the step size decreases adaptively with temperature during the course of annealing. Overall, we did not manage to achieve a faster decrease than $T^{-1}$ with these attempts. 

\section{Simulated annealing using a logarithmic schedule}
\label{sec.SA.log schedule}
\subsection{Logarithmic schedule}

In Ref.~\cite{Shinomoto91}, Shinomoto and Kabashima studied simulated annealing of the original form of $V_{\rm SK}(x)$, where the barrier height and distance between neighboring minima are kept constant for all $x$, under a logarithmic annealing schedule with time running from 0 to $\infty$ in contrast to the finite-time protocols in the previous sections. To test their prediction, let us consider the following schedule for the inverse temperature
\begin{equation}
\beta^{(l)}(t)= \beta_i \log_{10}\left( t + 10 \right),
\label{eq.log schedule.definition}
\end{equation}
where $\beta_i$ is the initial inverse temperature at time $t=0$. Time $t$ is supposed to run from 0 to $\infty$. The superscript $(l)$ denotes `logarithmic' and is to differentiate Eq. (\ref{eq.log schedule.definition}) from the other schedules. Shinomoto and Kabashima showed, by assuming that particle are located only at local minima and taking the continuum limit ignoring discreteness of positions, that the average of the potential energy during simulated annealing under Eq. (\ref{eq.log schedule.definition}) should decrease with time asymptotically as 
\begin{equation}
\langle V_{\rm SK}(x)\rangle_t \sim (\log_{10} t)^{-1},
\label{eq.SKs finite-time energy scaling}
\end{equation}
where $\langle V_{\rm SK}(x)\rangle_t$ is given by the right side of Eq. (\ref{eq.average potential energy.definition}), with a small difference that $\beta$ is now further dependent on time via the annealing schedule. We indicate this time dependence using the subscript $t$ in $\langle\cdot\rangle_t$. In the following, we shall study the dependence of $\langle V_{\rm SK}(x)\rangle_t$ on time from the perspective of Monte Carlo simulations to see if their prediction is observed numerically.

\subsection{Time scaling of the potential energy from Monte Carlo simulations}
\label{subsec.Finite-time scaling of the potential energy from Monte Carlo simulations}

\begin{figure}[h]
\begin{center}
\includegraphics[scale=0.75]{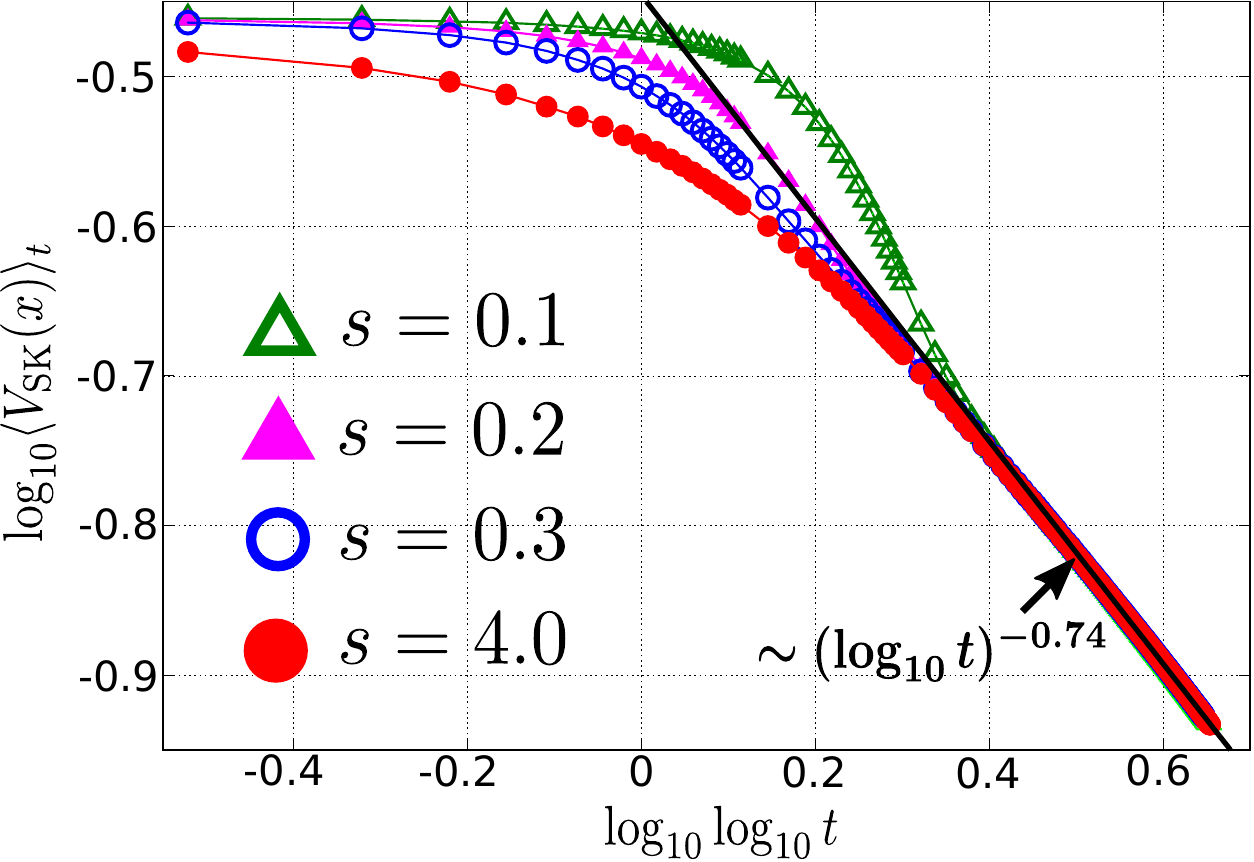}
\caption{Decay of $\langle V_{\rm SK}(x) \rangle_t$ with time under the schedule Eq. (\ref{eq.log schedule.definition}) ($\beta_i=10^{0.29}$). Results for several different $s$ are shown. Ensemble size is $N=10^7$ for all the curves. The errors due to finite-$N$ fluctuations are small and hence not shown. The curves collapse together asymptotically, decaying as $(\log_{10}t)^{-0.74}$ (solid line). Lines connecting the data points are to guide the eye only.}
\label{fig.average Vsk w time. various s}
\end{center}
\end{figure}

Figure \ref{fig.average Vsk w time. various s} shows the decay of $\langle V_{\rm SK}(x)\rangle_t$ with time under Eq. (\ref{eq.log schedule.definition}), with $\beta_i=10^{0.29}$. The initial ensemble at the start of annealing is the one drawn from the Boltzmann distribution of $V_{\rm SK}(x)$ with temperature $\beta_i$. Each curve in the figure represents the result obtained using a particular step size $s$. It is seen that when $t$ is large, the curves collapse onto a single curve. We fitted a straight line to the curve of $s=4$ using the data points from $\log_{10}\log_{10}t>0.5$, and the result is shown by the solid line (black). We see that $\langle V_{\rm SK}(x)\rangle_t$ decays asymptotically as $(\log_{10}t)^{-0.74}$. The decay exponent $-0.74$ is different from the $-1$ proposed by Shinomoto and Kabashima.

To account for our results, we propose generalizing the scaling Eq. (\ref{eq.SKs finite-time energy scaling}) slightly as 
\begin{equation}
\langle V_{\rm SK}(x)\rangle_t \sim (\log_{10} t)^{\alpha_{\rm s.a.}}
\label{eq.proposed new scaling}
\end{equation}
where the decay exponent $\alpha_{\rm s.a.}$ is obtained from numerical simulations. We repeated the simulated annealing calculations with different values of $\beta_i$ and the results are summarized in Table \ref{tab.summary of exponents.SA.log schedule}. These results capture the non-trivial aspects of the potential $V_{\rm SK}(x)$. The column $\beta_{\rm final}$ indicates the final inverse temperature attained by the schedule for that annealing simulation. By referring $\beta_{\rm final}$ to Fig. \ref{fig. E0 and U vs m and beta}(b), we see that our annealing was performed in the regime where the system exhibits `excess energy'.

\begin{table}[h]
\begin{center}
\begin{tabular}{c cc c cc c cc c}
\hline
\hline
     $\beta_i$ &&&   $\beta_{\rm final}$    &&& $\alpha_{\rm s.a.}$ &&& $\alpha_{\rm eq.}(\beta_{\rm final})$ \\
\hline
  $10^{0.29}$  &&& $10^{0.94}$   &&& $-0.74$ &&& $-0.74$\\
  $10^{1.00}$  &&& $10^{1.66}$   &&& $-1.07$ &&& $-1.07$ \\
  $10^{1.75}$  &&& $10^{2.40}$   &&& $-1.56$ &&& $-1.59$ \\
\hline
\hline
\end{tabular}
\caption{Decay exponent $\alpha_{\rm s.a.}$ obtained from Monte Carlo simulations. Annealing was performed under the schedule Eq. (\ref{eq.log schedule.definition}), starting from $\beta_i$ and ending at $\beta_{\rm final}$. The gradient $\alpha_{\rm eq.}(\beta_{\rm final})$ [c.f. Eq. (\ref{eq.local tangent.definition})] approximates $\alpha_{\rm s.a.}$ very well.}
\label{tab.summary of exponents.SA.log schedule}
\end{center}
\end{table}

The exponents $\alpha_{\rm s.a.}$ may be understood on the basis of the equilibrium properties of the system. Let us define
\begin{equation}
\alpha_{\rm eq.}(\beta')=\left. \frac{ d [\log_{10} \langle V_{\rm SK}(x)\rangle_{\beta}] }{d [\log_{10} \beta]} \right|_{\beta'}
\label{eq.local tangent.definition}
\end{equation}
where $\langle V_{\rm SK}(x)\rangle_{\beta}$ is given by Eq. (\ref{eq.average potential energy.definition}) and is the average potential energy at equilibrium. The quantity $\alpha_{\rm eq.}(\beta')$ is the gradient of the curve $\log_{10}\langle V_{\rm SK}(x)\rangle_{\beta}$ at $\log_{10}\beta'$. In Table \ref{tab.summary of exponents.SA.log schedule}, the column $\alpha_{\rm eq.}(\beta_{\rm final})$ shows the gradient evaluated at the final attained temperature $\beta_{\rm final}$, and we see that the values agree quite well with the decay exponent $\alpha_{\rm s.a.}$. This agreement between the two quantities can be explained by noting that when $t$ is large the schedule Eq. (\ref{eq.log schedule.definition}) varies very slowly. The system is therefore very close to equilibrium during the late stages of annealing, and it traces out a narrow segment of the $\langle V_{\rm SK}(x)\rangle_{\beta}$ curve around $\beta_{\rm final}$. The relation Eq. (\ref{eq.proposed new scaling}) can therefore be approximated by $\langle V_{\rm SK}(x)\rangle_{\beta_{\rm final}} \sim (\beta_{\rm final})^{\alpha_{\rm s.a.}}$, from which the interpretation of $\alpha_{\rm s.a.}$ as the gradient $\alpha_{\rm eq.}(\beta_{\rm final})$ follows immediately via Eq. (\ref{eq.local tangent.definition}).

We close this section with some comments on the decay exponent $-1$ obtained by Shinomoto and Kabashima and the $\alpha_{\rm s.a.}$ we obtained. One possible reason for the discrepancy is that the potential $V_{\rm SK}(x)$ we used, although inspired by Shinomoto and Kabashima's work, is not exactly the same as the one they treated. Another cause may be the coarse-graining assumption $(w_0)^2\beta_{\rm final}\ll 1$ which the authors made during their derivation. Substituting $w_0=0.2$ and the values of $\beta_{\rm final}$ in Table \ref{tab.summary of exponents.SA.log schedule}, we see that the inequality is not very well-satisfied for the numerical simulations which we performed here. 
A third, and possibly more important, reason may be as follows.  If the system is close to equilibrium and the particle stays almost exclusively in the valley around the global minimum at $x=0$ for $t$ large and thus $\beta$ large, then the equipartition law suggests that the average of the potential energy is proportional to $\beta^{-1}$, which would imply  $\langle V_{\rm SK}(x)\rangle_{\beta} \sim (\log_{10} t)^{-1}$ under the annealing schedule of Eq.~(\ref{eq.log schedule.definition}). Also the kinetic energy term leads to the same behavior due to equipartition. Deviations from this inverse-log law in numerical simulations would imply a few possibilities including a spreading of probability distribution away from the central valley to nearby local minima as well as deviations from equilibrium. It is anyway the case that the contribution to the equilibrium internal energy from kinetic energy  would yield the inverse-log law under the quasi-equilibrium context, which is more dominant than the potential term with larger powers as listed in Table \ref{tab.summary of exponents.SA.log schedule}.


\section{Summary and discussions}
\label{sec.Summary and discussions}

We have studied quantum and simulated annealing of a one-dimensional particle along the $x$-axis. The subject of optimization is played by the potential energy $V_{\rm SK}(x)$, which is modelled closely after a similar potential originally proposed by Shinomoto and Kabashima \cite{Shinomoto91}. On the one hand, the rugged energy landscape of $V_{\rm SK}(x)$ offers a non-trivial challenge for the annealing algorithms to solve; on the other, the potential is designed such that it possesses a unique global minimum at the origin. This combination of both difficulty and simplicity makes $V_{\rm SK}(x)$ a suitable candidate for understanding the fundamental aspects of quantum annealing for continuous variables, as well as for identifying the underlying differences between it and simulated annealing.

We first briefly reviewed the general features of $V_{\rm SK}(x)$ and its phase diagram. Focusing on a particular phase point, we then examined the properties of both the quantum ground state and the classical equilibrium state in detail. It was pointed out that the quantum ground state wavefunction appears more suitable for optimization than the classical Boltzmann distribution because it exhibits less signs of trapping by local minima at the same value of energy. We moved on to look at the dynamical aspects involving annealing. Quantum annealing was performed under linear and nonlinear schedules. The performance of annealing was assessed using the residual energy, which was found to decrease with total annealing time $T$ as $T^{-2n}$ where $n$ is the order of the schedule, which is in contrast to the result $T^{-1}$ by Stella {\em et al.} \cite{Stella05} derived by a phenomenological approach to the quantum process involving several steps of approximations. Also of particular interest is the use of probability current to monitor the time evolution of the wavefunction during the annealing process. The current revealed the different mechanisms at work behind linear and nonlinear schedules (adiabaticity versus persistent tunneling). We then went on to consider simulated annealing. Here, our main finding is that under the linear annealing schedule the residual energy decreases as $T^{-1}$ if we choose the step size $s$ appropriately. An important lesson we learnt is that the Monte Carlo step size utilized for annealing must be large enough to traverse the energy barrier between local minima to achieve the best performance; otherwise, the particle will be trapped and the residual energy will decrease much slower than $T^{-1}$, possibly logarithmically as predicted by Shinomoto and Kabashima.

The central result of this work is the comparison between the residual energies of quantum and simulated annealing. We have seen that under the linear schedule, the residual energy of quantum annealing decreases as $T^{-2}$ whereas that of simulated annealing decreases as $T^{-1}$ at best, i.e. even with quasi-global search processes across energy barriers, and probably as $(\log_{10} T)^{-1}$ under the usual local search. This provides another example in which quantum mechanics helps the system reach the global minimum more efficiently than classical stochastic process does. Furthermore, the theoretical framework of quantum annealing allows room for additional improvement, as we have seen from the results of nonlinear schedules. On the other hand, in the case of simulated annealing it is not immediately apparent how one can improve upon the scaling of $T^{-1}$ for quasi-global search and $(\log_{10} T)^{-1}$ for local search. Overall, our results seem to favor quantum over simulated annealing when it comes to the optimization of continuous variables, at least for potentials like $V_{\rm SK}(x)$.

A second point concerns the mechanisms underlying quantum annealing. We have seen that there are two different ways in which quantum annealing can arrive at the global minimum. Under the linear schedule, this goal is reached by suppressing the probability current so that the annealing mimics an adiabatic process. Under nonlinear schedules, on the other hand, the same objective is achieved by inducing a persistent tunneling current. These are two incompatible mechanisms, but interestingly both can be utilized by quantum annealing, depending on the way it is performed. In the literature, quantum adiabatic computation and quantum annealing are sometimes loosely used as synonyms. Our work shows that it is better not to conflate these two terms as the underlying mechanism can be very different.

Lastly, let us comment on possible ways to extend this work. One interesting feature of quantum annealing is the flexibility to choose (or even design) the annealing driver so as to reap the best performance. In this work, the role of the driver is played by the kinetic energy operator. One might consider introducing a secondary driver in the form of a vector potential operator $A(x)$, generalizing the momentum operator $p$ to $p-q(t)A(x)$. The time-dependent charge $q(t)$ plays the role of a second schedule, together with the mass $m(t)$. One recovers the original model when $q(t)=0$. It would be interesting to see if one could improve the annealing performance with some appropriate choice of $A(x)$ and $q(t)$. This could be useful, for instance, in Stage 2 where we have seen that resorting to nonlinear schedules did not deliver the expected improvement in the scaling of the residual energy.

\begin{acknowledgements}

This work is based on a project commissioned by the New Energy and Industrial Technology Development Organization.

\end{acknowledgements}

\appendix


\section{Overview of \texorpdfstring{$E_0(m)$}{Lg} and  \texorpdfstring{$\Delta(m)$}{Lg} in the  \texorpdfstring{$h_0$-$w_0$}{Lg} phase space}

\label{app. overview of E0 and gap in phase space}

In Sec. \ref{sec. Compare statics}, we presented a detailed discussion of the quantum ground state energy $E_0$ and the energy gap $\Delta$ of the system as a function of mass $m$. The discussion there focused on one point of $h_0$-$w_0$ phase space, $(w_0,h_0)=(0.2,0.2)$. In this appendix, we broaden the discussion to the curves of $E_0(m)$ and $\Delta(m)$ at other phase points, with the aim of providing a more comprehensive picture of the time-independent aspects of the system.

\begin{figure}[h]
\begin{center}
\includegraphics[scale=0.65]{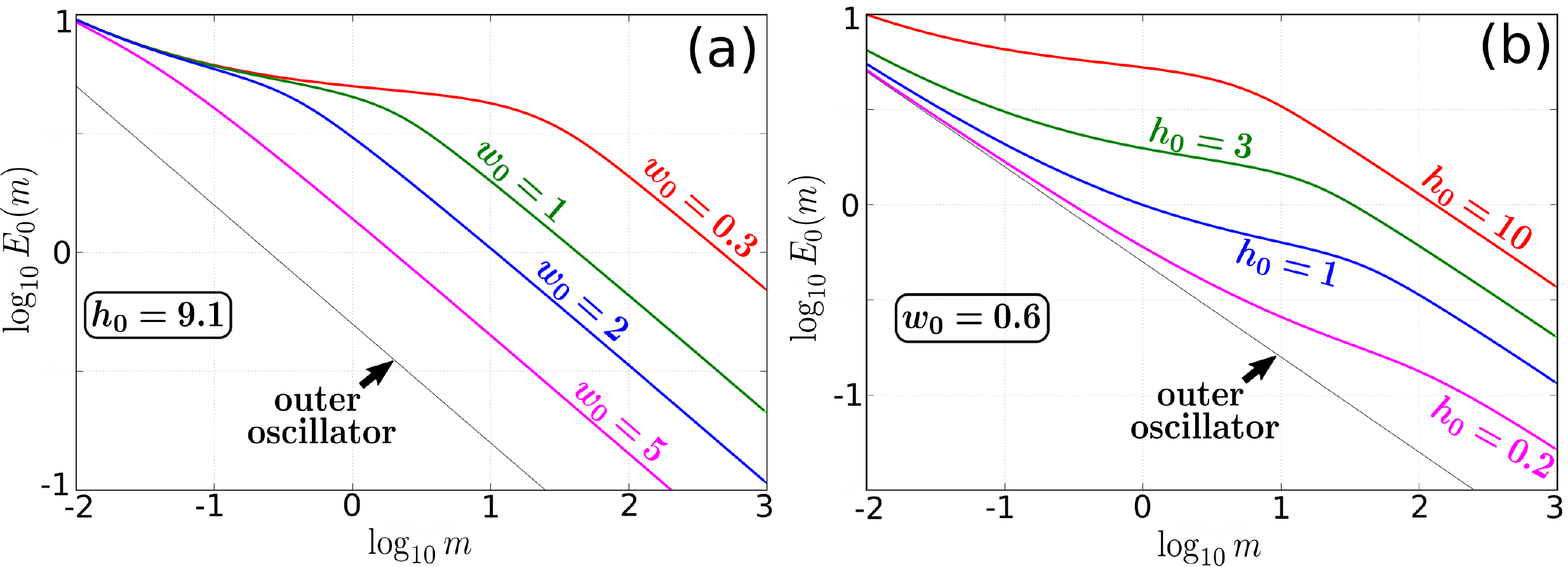}
\caption{Ground state energy $E_0$ as a function of mass $m$, evaluated at points in $h_0$-$w_0$ phase space other than $(0.2,0.2)$. The lines labelled `outer oscillator' have the same meaning as in Fig.~\ref{fig. E0 and U vs m and beta}. (a) The $E_0(m)$ curves where $h_0=9.1$ and $w_0$ varies. (b) The curves where $w_0=0.6$ and $h_0$ varies.}
\label{fig.app.E0(m).fixed h0 and w0}
\end{center}
\end{figure}

We first discuss the energy $E_0$. As seen in Fig. \ref{fig. E0 and U vs m and beta}(a), the curve of $E_0(m)$ has an `S' shape, resembling the function $\tanh x$. This feature is generally true at other phase points as well. Panel (a) of Fig. \ref{fig.app.E0(m).fixed h0 and w0} shows several $E_0(m)$ curves where $h_0$ is fixed and $w_0$ varies. [For $w_0=5$ (magenta), the `S' portion of the curve lies in the range $m<10^{-2}$.] Panel (b) shows the complementary situation where $w_0$ is fixed and $h_0$ varies. Qualitatively, we see that the energy curve of Fig. \ref{fig. E0 and U vs m and beta}(a) at the phase point $(0.2,0.2)$ can be considered a reasonable representative of the generic behavior in phase space. Quantitatively, however, one also sees that the `S' transition at different phase points varies over a wide range of masses. Hence, one should be slightly cautious when inferring the behavior at other phase points based on the results obtained at $(0.2,0.2)$.

The energy gap $\Delta$, on the other hand, exhibits richer behavior compared to $E_0$. In addition to the `flat gap' described in Fig. \ref{fig. gap vs m }, we also observed two other notable features on the $\Delta(m)$ curves of other phase points.

\begin{figure}[h]
\begin{center}
\includegraphics[scale=0.65]{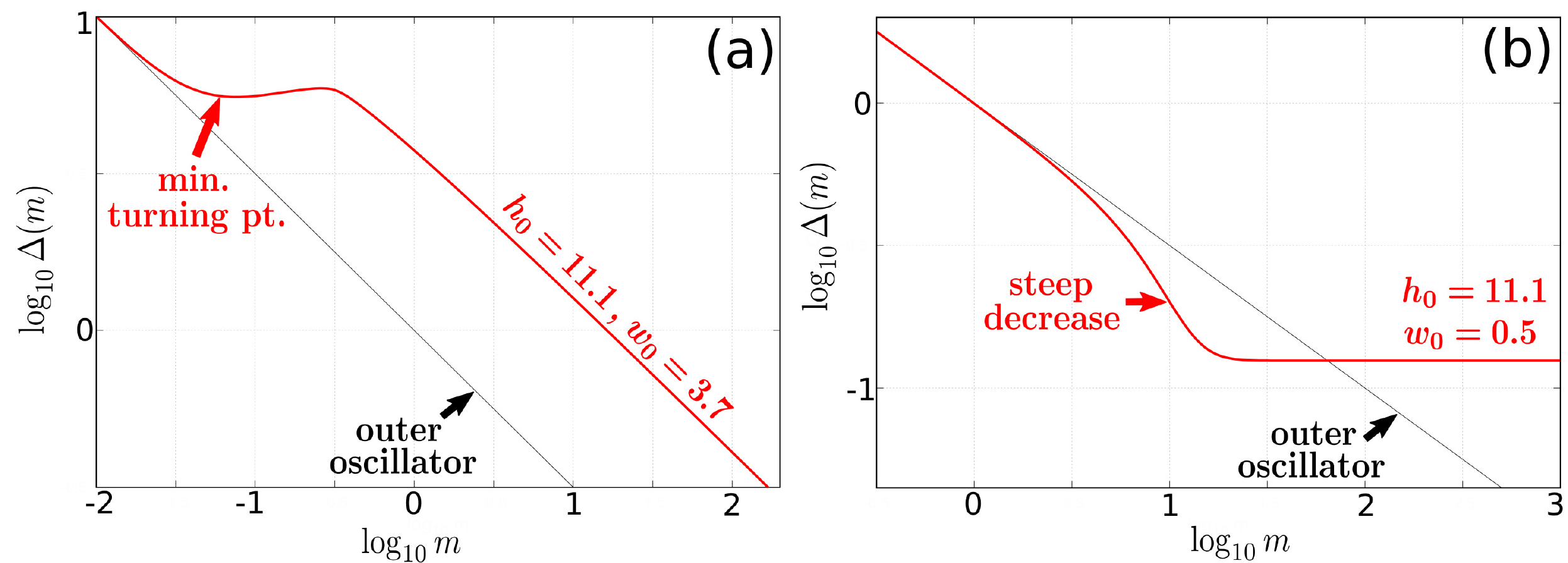}
\caption{Two notable features on the energy gap curves. Solid curves (red) show the energy gap $\Delta$ as a function of mass $m$. The lines labelled `outer oscillator' have the same meaning as in Fig. \ref{fig. gap vs m }. (a) The $\Delta(m)$ curve at the phase point $(w_0,h_0)=(3.7,11.1)$ has a minimum turning point when the mass is small. (b) At the phase point $(0.5,11.1)$, the gap of $V_{\rm SK}(x)$ shows a steep and significant decrease relative to the gap of the oscillator.}
\label{fig.app.Delta(m).two features}
\end{center}
\end{figure}

The first concerns a minimum turning point on the gap curve. In studies on quantum annealing, much attention is being paid to the minimum energy gap as it constitutes the bottleneck of the entire annealing process. For the potential $V_{\rm SK}(x)$, however, a minimum gap in the conventional sense does not exist. This is most easily seen by considering the gap of the harmonic oscillator, $\sqrt{k/m}$, which approaches zero as the annealing approaches completion at infinite mass. However, there are situations where the gap goes through a minimum turning point. Panel (a) of Fig. \ref{fig.app.Delta(m).two features} shows an example of such a gap curve obtained at the phase point $(w_0,h_0)=(3.7,11.1)$. Based on our observations of extensive numerical data, we found that (i) the turning point generally occurs at small $m$, and (ii) just the northeastern region of phase space (c.f. Fig. \ref{fig.N_min}) exhibits this feature on the gap curve. [The phase point $(0.2,0.2)$ lies in the southwestern region and does not exhibit a turning point.] In this paper, we did not investigate whether this minimum turning point has any influence on quantum annealing.

The second feature concerns the occurrence of small gaps. Once again, this is conventionally characterized by the minimum gap. In the case of $V_{\rm SK}(x)$, however, we found that the gap can show a significant decrease in magnitude but does not go through a turning point. Panel (b) of Fig. \ref{fig.app.Delta(m).two features} shows an example of such a gap curve obtained at the phase point $(w_0,h_0)=(0.5,11.1)$. The sudden and anomalous decrease in the gap size of $V_{\rm SK}(x)$ can be evinced when compared to the `benign' gap behavior of the oscillator. Note that the scale along the vertical axis is logarithmic, which means that the reduction in magnitude is quite significant. The effects of this peculiar gap feature on annealing is also not covered by our numerical studies.

\begin{figure}[h]
\begin{center}
\includegraphics[scale=0.65]{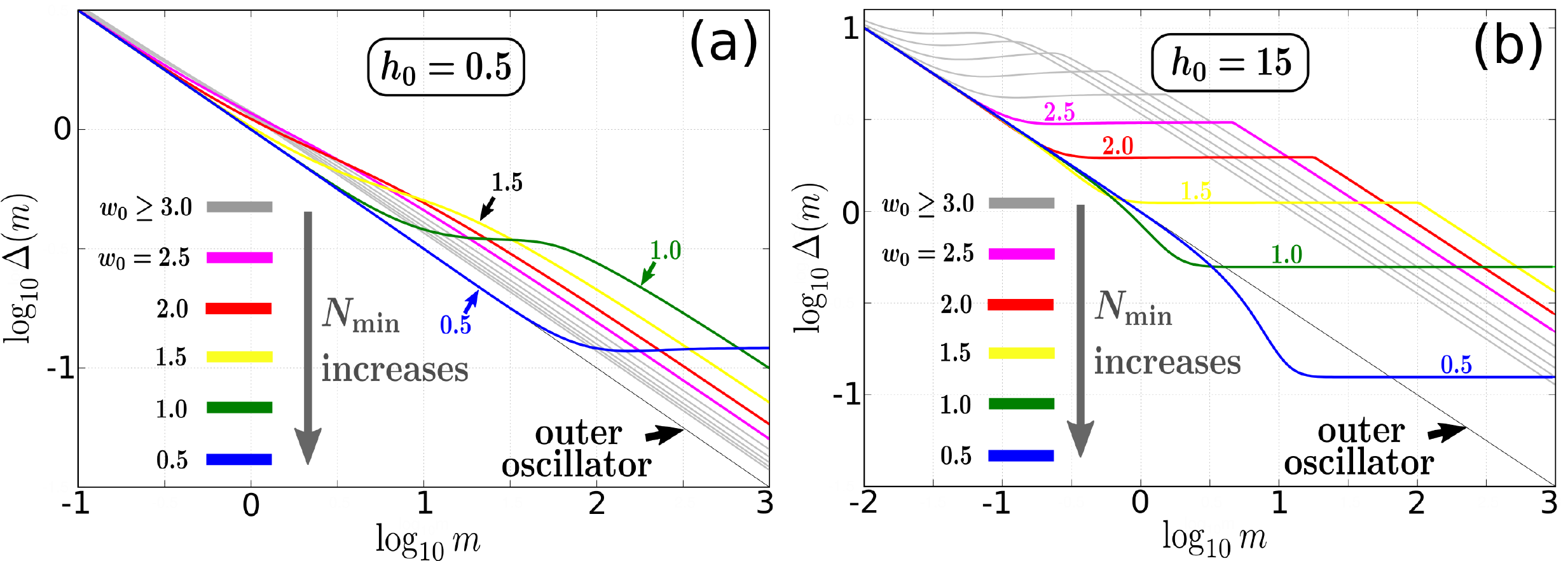}
\caption{Evolution of the gap curve as $w_0$ changes with $h_0$ held fixed. The curves exhibit salient changes when $w_0$ decreases below 2.5 and these are highlighted in color. To aid visualization, the other curves ($w_0\ge 3$) are plotted in gray. Note that decrease in $w_0$ corresponds to increase in $N_{\rm min}$ on the phase diagram, and hence  increased complexity of $V_{\rm SK}(x)$'s energy landscape. (a) For $h_0=0.5$ (small barrier).  (b) For $h_0=15$ (large barrier).}
\label{fig.app.Delta(m).fixed h0}
\end{center}
\end{figure}

The behavior of the $\Delta(m)$ curve at different points in phase space can be quite complex, exhibiting interplay among the various features mentioned above. Figure \ref{fig.app.Delta(m).fixed h0} shows some examples of how the curve evolves when $h_0$ is kept constant and $w_0$ varies. Panel (a) shows the case of $h_0=0.5$, corresponding to a small barrier in the potential $V_{\rm SK}(x)$. The inter-minima distance $w_0$ is decreased from 5 to 0.5, which in phase space translates to an increasing $N_{\rm min}$. To aid visualization, only the curves showing notable changes are plotted with color. We see that the energy gap evolves from a simple, monotonically decreasing curve (gray) to one exhibiting a distinctive `flat gap' (blue). Panel (b) shows the case of large barrier $h_0=15$. The evolution of the gap curve here is more complicated, first showing a minimum turning point (gray), then a `flat gap' (red), and finally a `steep decrease' (blue). Based on extensive numerical observations, we found that in general when $N_{\rm min}$ increases beyond a certain threshold, the two features, `flat gap' followed by `steep decrease', will emerge on the $\Delta(m)$ curve.

\begin{figure}[h]
\begin{center}
\includegraphics[scale=0.65]{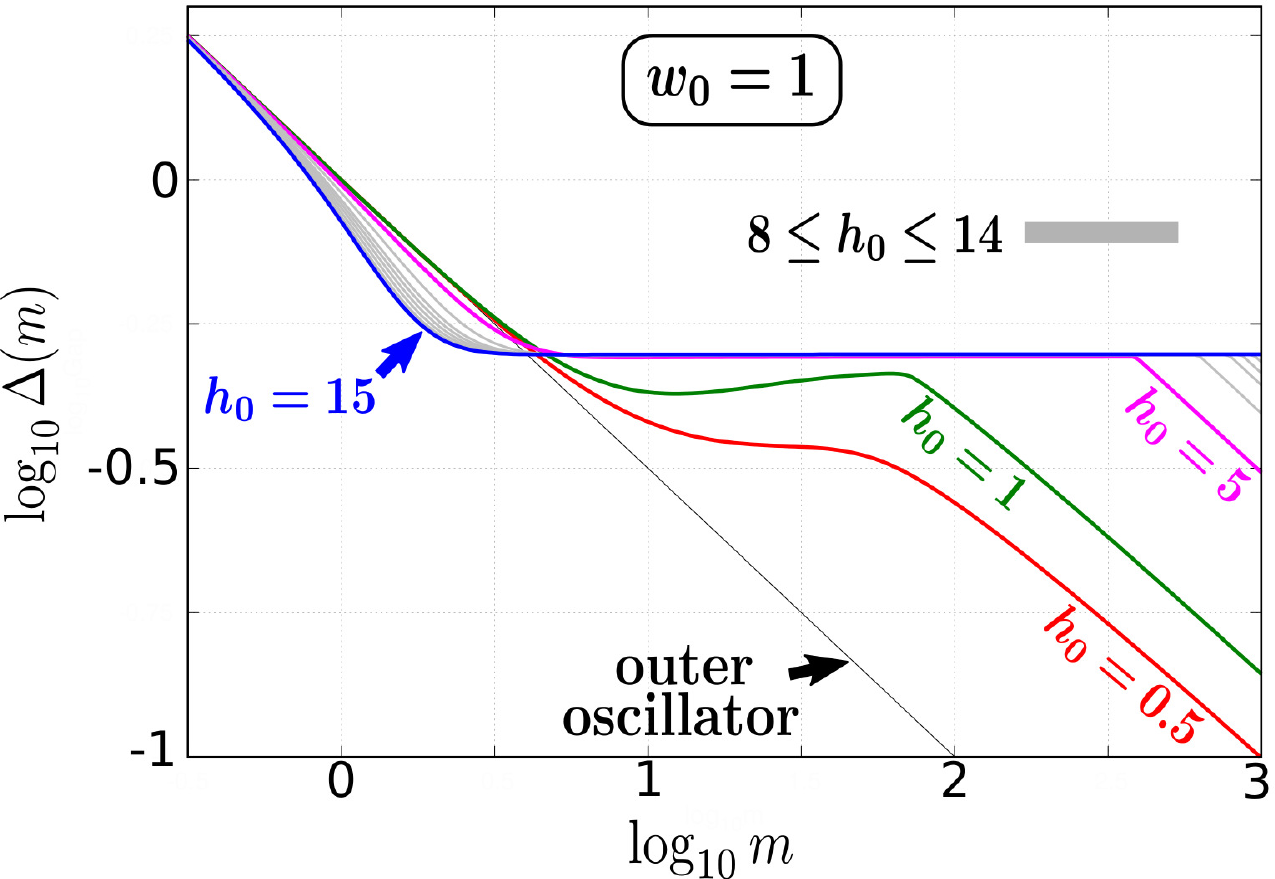}
\caption{Evolution of the gap curve when $w_0=1$ and $h_0$ increases from 0.5 to 15. To aid visualization, only the curves exhibiting prominent changes are plotted in color. Notice that for the gray curves ($8\le w_0\le 14$), their `flat gap' region (c.f. Fig. \ref{fig. gap vs m }) are collapsed onto a single horizontal line.}
\label{fig.app.Delta(m).fixed w0}
\end{center}
\end{figure}

Figure \ref{fig.app.Delta(m).fixed w0} shows the complementary situation, with $w_0$ held constant and $h_0$ varying. It is seen that as $h_0$ (and $N_{\rm min}$) increases, the gap curve evolves from being monotonically decreasing (red) to having a minimum turning point (green), and then finally attaining the features `flat gap' and `steep decrease' (blue). An additional point to note here is that for the gray curves, their `flat gap' segment seems to have collapsed onto a single horizontal line. This phenomenon is also a motif that recurs quite frequently throughout phase space.

To summarize, in this appendix we presented an overview of the behaviors of $E_0(m)$ and $\Delta(m)$ at different points in phase space. In general, the $E_0(m)$ curve evolves quite predictably with respect to changes in the parameters $h_0$ and $w_0$. By contrast, the shape of the $\Delta(m)$ curve is more diversed, exhibiting different features at different points in phase space. The discussion here aims to give the reader a sense of the underlying complexity of $V_{\rm SK}(x)$, while also serving as a caveat that our annealing studies performed at just one phase point might not be exhaustive in terms of capturing the dynamical aspects of the system.

\section{Time evolution of probability current in Stages 2 and 3}
\label{app.time evolution of current in stages 2 and 3}

In Secs. \ref{subsec.time evolution.QA.linear} and \ref{subsec.persistent tunneling under nonlinear schedules}, we discussed the temporal behavior of the probability current $j(x,t)$ in Stage 1, and it was mentioned that the results there are also valid in Stages 2 and 3. In this Appendix, we supplement those discussions by presenting some of the key results in these two stages.

\subsection{Stage 2 under the linear schedule}

The probability current in Stage 2 is sometimes not spatially-coherent, meaning that the current flows in different directions at different points in space. Figure \ref{fig.snapshot of j(x,t).stage2.linear.t=10.T=320} shows a snapshot of $j(x,t)$ in this stage under the linear schedule at time $t=10$, with annealing time $T=320$. It is seen that $j(x,t)$ undulates between positive and negative values spatially, so we cannot represent its overall spatial behavior with just the amplitude $j(x_0,t)$ as in Fig. \ref{fig.explaining width and amplitude}(b). To keep things simple, let us just consider $j(x,t)$ at three locations, namely $x=-0.025,$ $-0.1$, and $-0.175$, indicated by the vertical lines in Fig. \ref{fig.snapshot of j(x,t).stage2.linear.t=10.T=320}. Physically speaking, $x=-0.1$ is located at the first local maximum of $V_{\rm SK}(x)$ and monitors the current in the classically forbidden region; the other two points monitor the currents in the vicinity of the global minimum and its adjacent local minimum.

\begin{figure}[h]
\begin{center}
\includegraphics[scale=0.80]{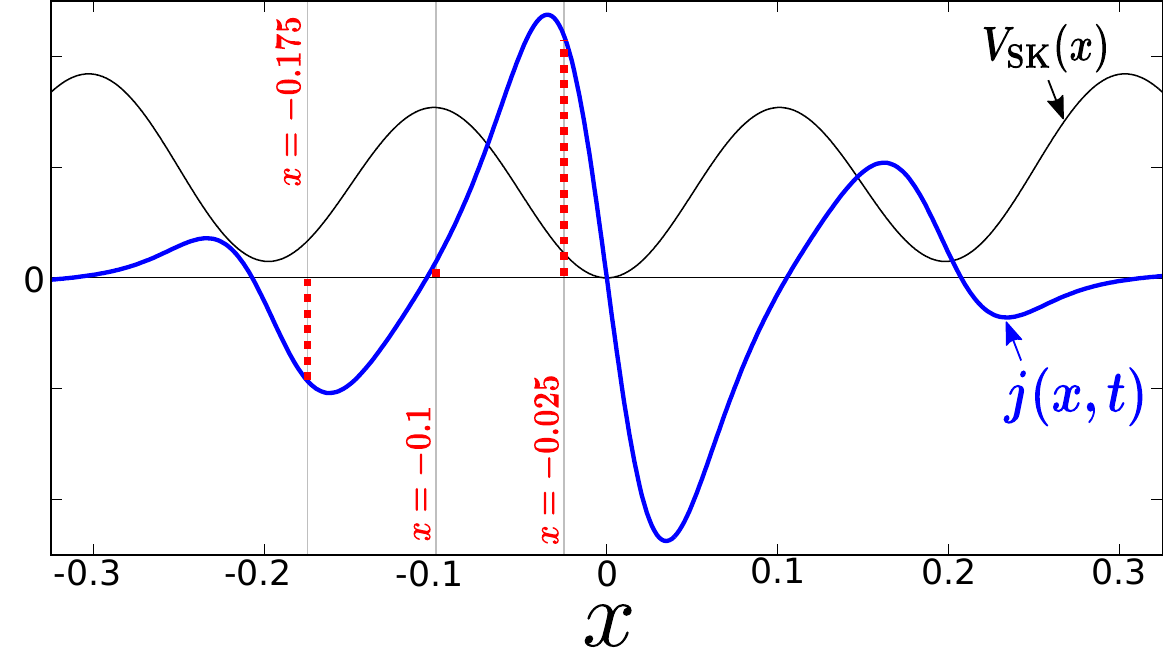}
\caption{Snapshot of the probability current $j(x,t)$ at time $t=10$ in Stage 2 under the linear schedule with $T=320$. The undulation of $j(x,t)$ between positive and negative values spatially indicates that the current flows in different directions at different points in space. We monitor the time evolution of $j(x,t)$ at $x=-0.025, -0.1$ and $-0.175$ (vertical lines), three $x$ positions located between the global minimum and the first local minimum of $V_{\rm SK}(x)$.}
\label{fig.snapshot of j(x,t).stage2.linear.t=10.T=320}
\end{center}
\end{figure}

The time evolutions are presented in Fig. \ref{fig.compare j(x,t).3 x pts.Stage2.linear.3Ts}, where panels (a) to (c) show the results for $T=320$, 3250, and 10000, respectively. In each panel, the current $j(x,t)$ at the three $x$ positions are plotted as a function of time. Generally speaking, we see that as $T$ increases, the oscillations of $j(x,t)$ at all three $x$ positions become suppressed and the annealing approaches adiabaticity. This is similar to what we have seen in Stage 1 for the linear schedule [c.f. Sec. \ref{subsec.time evolution.QA.linear} and Fig. \ref{fig.time evoluton.width and amplitude.stage1.linear.various Ts}(b)].

\begin{figure}[h]
\begin{center}
\includegraphics[scale=0.80]{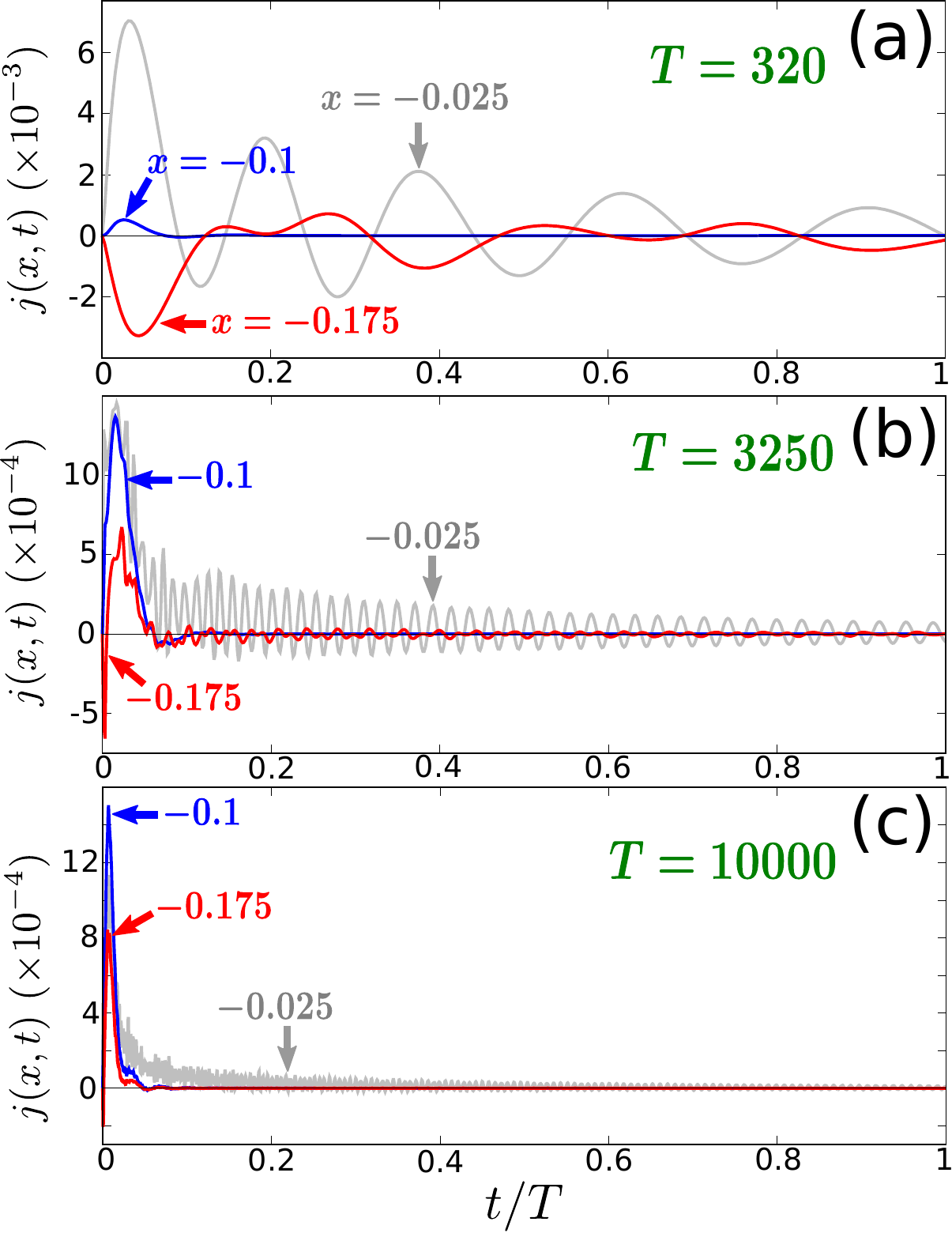}
\caption{Time evolutions of $j(x,t)$ at $x=-0.025, -0.1,$ and $-0.175$ (c.f. Fig. \ref{fig.snapshot of j(x,t).stage2.linear.t=10.T=320}) in Stage 2 under the linear schedule. For total annealing times (a) $T=320$, (b) $T=3250$, and (c) $T=10000$. Generally, the behaviors at all three spatial points exhibit the same trend towards adiabaticity as in Stage 1 [c.f. Fig. \ref{fig.time evoluton.width and amplitude.stage1.linear.various Ts}(b)].}
\label{fig.compare j(x,t).3 x pts.Stage2.linear.3Ts}
\end{center}
\end{figure}

Let us make a minor point concerning the current near the local minimum ($x=-0.175$, i.e. the red curves). When the total time is short ($T=320$), during the early stages the current is negative, so it is directed towards the local minimum. This explains the spurious peaks observed in Fig. \ref{fig.badly annealed wavefunction.Stage 2 linear schedule T=3250}. As $T$ increases, this early current becomes predominantly positive, meaning that it now flows away from the local minimum and towards the global one by tunneling, thereby aiding annealing in attaining the correct optimized solution. Nevertheless, we also see that the magnitude and duration of this reversed current decrease with longer $T$, so adiabaticity is ultimately still the main mechanism contributing to a successful optimization here.

\subsection{Stage 2 under nonlinear schedules}

In Sec. \ref{subsec.persistent tunneling under nonlinear schedules}, we have seen that under nonlinear schedules the probability current in Stage 1 exhibits persistent tunneling, which is in contrast to the oscillatory behavior seen under the linear schedule. The corresponding situation in Stage 2 is quite similar. First, let us note that under nonlinear schedules, the current $j(x,t)$ in Stage 2 has the spatially-coherent form shown in Fig. \ref{fig.explaining width and amplitude}(b), so the amplitude $j(x_0,t)$ can again be used as a parameter for monitoring the temporal behavior. Figure \ref{fig.compare j(x0,t) nonlinear schedules.stage2.T3250} shows the time evolutions of the amplitude in Stage 2 under the various nonlinear schedules, with $T=3250$. Overall, the behaviors are similar to what we have seen in Stage 1 (c.f. Fig. \ref{fig.compare  j(x0,t) various schedules.stage1.T560}).

\begin{figure}[h]
\begin{center}
\includegraphics[scale=0.9]{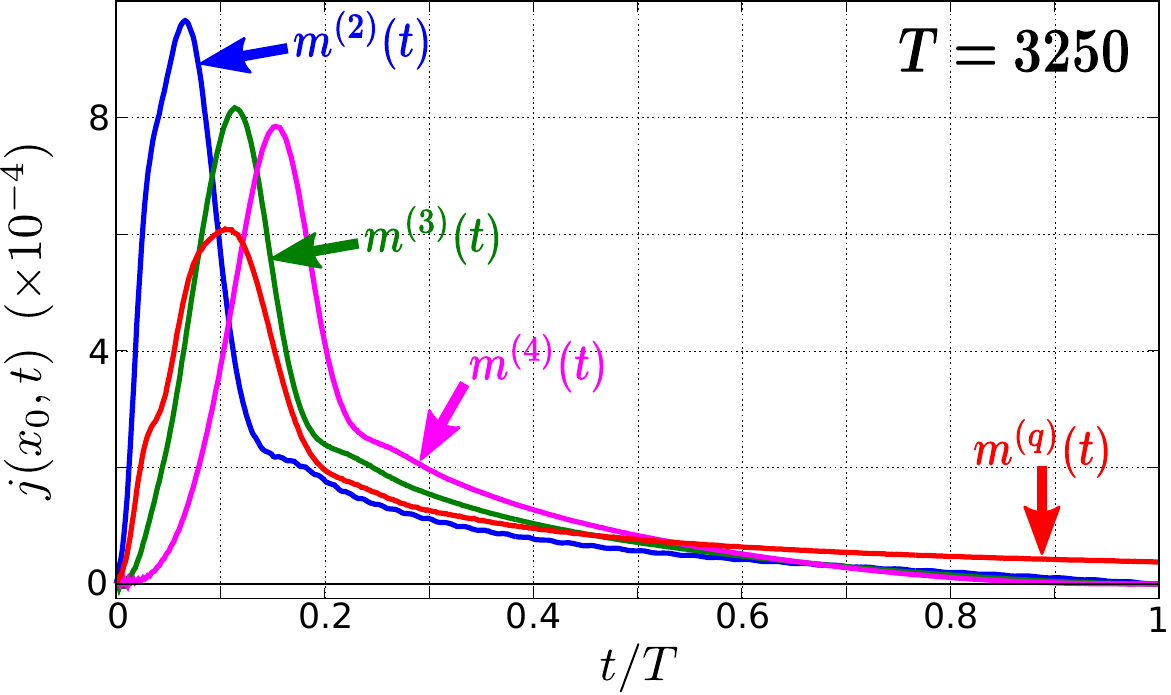}
\caption{Comparing the time evolutions of the current amplitude $j(x_0,t)$ in Stage 2 under nonlinear schedules. In general, the amplitude exhibits similar behavior as in Stage 1 (c.f. Fig. \ref{fig.compare  j(x0,t) various schedules.stage1.T560}), showing no oscillations and staying strictly positive throughout the entire annealing.}
\label{fig.compare j(x0,t) nonlinear schedules.stage2.T3250}
\end{center}
\end{figure}

\subsection{Stage 3 under the linear schedule}

In Stage 3, the wavefunction is well-localized within the central global minimum [c.f. Fig. \ref{fig. compare quantum classical groundstates . 0.1 and 0.01}(c)], and so annealing can be thought of as taking place within an effective quadratic potential. Figure \ref{fig.j(x0,t) stage3.linear.T1000} shows the time evolution of the amplitude $j(x_0,t)$ in Stage 3 under the linear schedule, with $T=1000$. We see that the current exhibits the oscillatory behavior seen in Stages 1 and 2. A minor observation would be that the crests (positive current) are larger in magnitude than the troughs (negative current), so the wavefunction becomes narrower after each oscillatory cycle.

For nonlinear schedules, the overall behavior is similar to Stages 1 and 2, so we shall not discuss them further.

\begin{figure}[h]
\begin{center}
\includegraphics[scale=1]{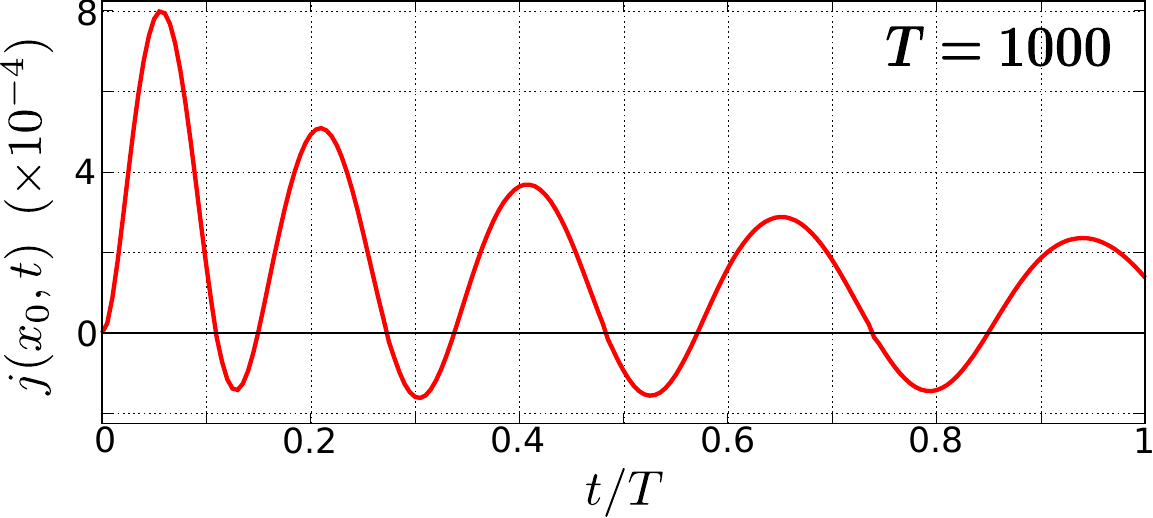}
\caption{Time evolution of the current amplitude $j(x_0,t)$ in Stage 3 under the linear schedule.}
\label{fig.j(x0,t) stage3.linear.T1000}
\end{center}
\end{figure}

\input{refs.bbl}

\end{document}

%% file: refs.bbl
%

%% file: manuscript_v2.bbl
\begin{thebibliography}{25}%
\makeatletter
\providecommand \@ifxundefined [1]{%
 \@ifx{#1\undefined}
}%
\providecommand \@ifnum [1]{%
 \ifnum #1\expandafter \@firstoftwo
 \else \expandafter \@secondoftwo
 \fi
}%
\providecommand \@ifx [1]{%
 \ifx #1\expandafter \@firstoftwo
 \else \expandafter \@secondoftwo
 \fi
}%
\providecommand \natexlab [1]{#1}%
\providecommand \enquote  [1]{``#1''}%
\providecommand \bibnamefont  [1]{#1}%
\providecommand \bibfnamefont [1]{#1}%
\providecommand \citenamefont [1]{#1}%
\providecommand \href@noop [0]{\@secondoftwo}%
\providecommand \href [0]{\begingroup \@sanitize@url \@href}%
\providecommand \@href[1]{\@@startlink{#1}\@@href}%
\providecommand \@@href[1]{\endgroup#1\@@endlink}%
\providecommand \@sanitize@url [0]{\catcode `\\12\catcode `\$12\catcode
  `\&12\catcode `\#12\catcode `\^12\catcode `\_12\catcode `\%12\relax}%
\providecommand \@@startlink[1]{}%
\providecommand \@@endlink[0]{}%
\providecommand \url  [0]{\begingroup\@sanitize@url \@url }%
\providecommand \@url [1]{\endgroup\@href {#1}{\urlprefix }}%
\providecommand \urlprefix  [0]{URL }%
\providecommand \Eprint [0]{\href }%
\providecommand \doibase [0]{http://dx.doi.org/}%
\providecommand \selectlanguage [0]{\@gobble}%
\providecommand \bibinfo  [0]{\@secondoftwo}%
\providecommand \bibfield  [0]{\@secondoftwo}%
\providecommand \translation [1]{[#1]}%
\providecommand \BibitemOpen [0]{}%
\providecommand \bibitemStop [0]{}%
\providecommand \bibitemNoStop [0]{.\EOS\space}%
\providecommand \EOS [0]{\spacefactor3000\relax}%
\providecommand \BibitemShut  [1]{\csname bibitem#1\endcsname}%
\let\auto@bib@innerbib\@empty
\bibitem [{\citenamefont {Kadowaki}\ and\ \citenamefont
  {Nishimori}(1998)}]{Kadowaki1998}%
  \BibitemOpen
  \bibfield  {author} {\bibinfo {author} {\bibfnamefont {T.}~\bibnamefont
  {Kadowaki}}\ and\ \bibinfo {author} {\bibfnamefont {H.}~\bibnamefont
  {Nishimori}},\ }\href {\doibase 10.1103/PhysRevE.58.5355} {\bibfield
  {journal} {\bibinfo  {journal} {Phys. Rev. E}\ }\textbf {\bibinfo {volume}
  {58}},\ \bibinfo {pages} {5355} (\bibinfo {year} {1998})}\BibitemShut
  {NoStop}%
\bibitem [{\citenamefont {Santoro}\ \emph {et~al.}(2002)\citenamefont
  {Santoro}, \citenamefont {Marto{\v n}{\'{a}}k}, \citenamefont {Tosatti},\
  and\ \citenamefont {Car}}]{Santoro2002}%
  \BibitemOpen
  \bibfield  {author} {\bibinfo {author} {\bibfnamefont {G.~E.}\ \bibnamefont
  {Santoro}}, \bibinfo {author} {\bibfnamefont {R.}~\bibnamefont {Marto{\v
  n}{\'{a}}k}}, \bibinfo {author} {\bibfnamefont {E.}~\bibnamefont {Tosatti}},
  \ and\ \bibinfo {author} {\bibfnamefont {R.}~\bibnamefont {Car}},\ }\href
  {\doibase 10.1126/science.1068774} {\bibfield  {journal} {\bibinfo  {journal}
  {Science}\ }\textbf {\bibinfo {volume} {295}},\ \bibinfo {pages} {2427}
  (\bibinfo {year} {2002})}\BibitemShut {NoStop}%
\bibitem [{\citenamefont {Das}\ and\ \citenamefont
  {Chakrabarti}(2008)}]{Das2008}%
  \BibitemOpen
  \bibfield  {author} {\bibinfo {author} {\bibfnamefont {A.}~\bibnamefont
  {Das}}\ and\ \bibinfo {author} {\bibfnamefont {B.~K.}\ \bibnamefont
  {Chakrabarti}},\ }\href {\doibase 10.1103/RevModPhys.80.1061} {\bibfield
  {journal} {\bibinfo  {journal} {Rev. Mod. Phys.}\ }\textbf {\bibinfo {volume}
  {80}},\ \bibinfo {pages} {1061} (\bibinfo {year} {2008})}\BibitemShut
  {NoStop}%
\bibitem [{\citenamefont {Santoro}\ and\ \citenamefont
  {Tosatti}(2006)}]{Santoro:2006}%
  \BibitemOpen
  \bibfield  {author} {\bibinfo {author} {\bibfnamefont {G.~E.}\ \bibnamefont
  {Santoro}}\ and\ \bibinfo {author} {\bibfnamefont {E.}~\bibnamefont
  {Tosatti}},\ }\href {\doibase doi:10.1088/0305-4470/39/36/R01} {\bibfield
  {journal} {\bibinfo  {journal} {J. Phys. A}\ }\textbf {\bibinfo {volume}
  {39}},\ \bibinfo {pages} {R393} (\bibinfo {year} {2006})}\BibitemShut
  {NoStop}%
\bibitem [{\citenamefont {Morita}\ and\ \citenamefont
  {Nishimori}(2008)}]{Morita2008}%
  \BibitemOpen
  \bibfield  {author} {\bibinfo {author} {\bibfnamefont {S.}~\bibnamefont
  {Morita}}\ and\ \bibinfo {author} {\bibfnamefont {H.}~\bibnamefont
  {Nishimori}},\ }\href {\doibase 10.1063/1.2995837} {\bibfield  {journal}
  {\bibinfo  {journal} {J. Math. Phys.}\ }\textbf {\bibinfo {volume} {49}},\
  \bibinfo {pages} {125210} (\bibinfo {year} {2008})}\BibitemShut {NoStop}%
\bibitem [{\citenamefont {Albash}\ and\ \citenamefont
  {Lidar}(2018)}]{Albash2018}%
  \BibitemOpen
  \bibfield  {author} {\bibinfo {author} {\bibfnamefont {T.}~\bibnamefont
  {Albash}}\ and\ \bibinfo {author} {\bibfnamefont {D.~A.}\ \bibnamefont
  {Lidar}},\ }\href {\doibase 10.1103/RevModPhys.90.015002} {\bibfield
  {journal} {\bibinfo  {journal} {Rev. Mod. Phys.}\ }\textbf {\bibinfo {volume}
  {90}},\ \bibinfo {pages} {015002} (\bibinfo {year} {2018})}\BibitemShut
  {NoStop}%
\bibitem [{\citenamefont {Hauke}\ \emph {et~al.}(2020)\citenamefont {Hauke},
  \citenamefont {Katzgraber}, \citenamefont {Lechner}, \citenamefont
  {Nishimori},\ and\ \citenamefont {Oliver}}]{Hauke2020}%
  \BibitemOpen
  \bibfield  {author} {\bibinfo {author} {\bibfnamefont {P.}~\bibnamefont
  {Hauke}}, \bibinfo {author} {\bibfnamefont {H.~G.}\ \bibnamefont
  {Katzgraber}}, \bibinfo {author} {\bibfnamefont {W.}~\bibnamefont {Lechner}},
  \bibinfo {author} {\bibfnamefont {H.}~\bibnamefont {Nishimori}}, \ and\
  \bibinfo {author} {\bibfnamefont {W.~D.}\ \bibnamefont {Oliver}},\ }\href
  {https://doi.org/10.1088%2F1361-6633%2Fab85b8} {\bibfield  {journal}
  {\bibinfo  {journal} {Rep. Prog. Phys.}\ }\textbf {\bibinfo {volume} {83}},\
  \bibinfo {pages} {054401} (\bibinfo {year} {2020})}\BibitemShut {NoStop}%
\bibitem [{\citenamefont {Farhi}\ \emph {et~al.}(2001)\citenamefont {Farhi},
  \citenamefont {Goldstone}, \citenamefont {Gutmann}, \citenamefont {Lapan},
  \citenamefont {Lundgren},\ and\ \citenamefont {Preda}}]{Farhi2001}%
  \BibitemOpen
  \bibfield  {author} {\bibinfo {author} {\bibfnamefont {E.}~\bibnamefont
  {Farhi}}, \bibinfo {author} {\bibfnamefont {J.}~\bibnamefont {Goldstone}},
  \bibinfo {author} {\bibfnamefont {S.}~\bibnamefont {Gutmann}}, \bibinfo
  {author} {\bibfnamefont {J.}~\bibnamefont {Lapan}}, \bibinfo {author}
  {\bibfnamefont {A.}~\bibnamefont {Lundgren}}, \ and\ \bibinfo {author}
  {\bibfnamefont {D.}~\bibnamefont {Preda}},\ }\href {\doibase
  10.1126/science.1057726} {\bibfield  {journal} {\bibinfo  {journal}
  {Science}\ }\textbf {\bibinfo {volume} {292}},\ \bibinfo {pages} {472}
  (\bibinfo {year} {2001})}\BibitemShut {NoStop}%
\bibitem [{\citenamefont {Finnila}\ \emph {et~al.}(1994)\citenamefont
  {Finnila}, \citenamefont {Gomez}, \citenamefont {Sebenik}, \citenamefont
  {Stenson},\ and\ \citenamefont {Doll}}]{Finnila1994}%
  \BibitemOpen
  \bibfield  {author} {\bibinfo {author} {\bibfnamefont {A.~B.}\ \bibnamefont
  {Finnila}}, \bibinfo {author} {\bibfnamefont {M.~A.}\ \bibnamefont {Gomez}},
  \bibinfo {author} {\bibfnamefont {C.}~\bibnamefont {Sebenik}}, \bibinfo
  {author} {\bibfnamefont {C.}~\bibnamefont {Stenson}}, \ and\ \bibinfo
  {author} {\bibfnamefont {J.~D.}\ \bibnamefont {Doll}},\ }\href {\doibase
  10.1016/0009-2614(94)00117-0} {\bibfield  {journal} {\bibinfo  {journal}
  {Chemical Physics Letters}\ }\textbf {\bibinfo {volume} {219}},\ \bibinfo
  {pages} {343} (\bibinfo {year} {1994})}\BibitemShut {NoStop}%
\bibitem [{\citenamefont {Abel}\ and\ \citenamefont
  {Spannowsky}(2021)}]{Abel2021PRXQ}%
  \BibitemOpen
  \bibfield  {author} {\bibinfo {author} {\bibfnamefont {S.}~\bibnamefont
  {Abel}}\ and\ \bibinfo {author} {\bibfnamefont {M.}~\bibnamefont
  {Spannowsky}},\ }\href {\doibase 10.1103/PRXQuantum.2.010349} {\bibfield
  {journal} {\bibinfo  {journal} {PRX Quantum}\ }\textbf {\bibinfo {volume}
  {2}},\ \bibinfo {pages} {010349} (\bibinfo {year} {2021})}\BibitemShut
  {NoStop}%
\bibitem [{\citenamefont {Abel}\ \emph
  {et~al.}(2021{\natexlab{a}})\citenamefont {Abel}, \citenamefont
  {Chancellor},\ and\ \citenamefont {Spannowsky}}]{Abel2021PRD}%
  \BibitemOpen
  \bibfield  {author} {\bibinfo {author} {\bibfnamefont {S.}~\bibnamefont
  {Abel}}, \bibinfo {author} {\bibfnamefont {N.}~\bibnamefont {Chancellor}}, \
  and\ \bibinfo {author} {\bibfnamefont {M.}~\bibnamefont {Spannowsky}},\
  }\href {\doibase 10.1103/PhysRevD.103.016008} {\bibfield  {journal} {\bibinfo
   {journal} {Phys. Rev. D}\ }\textbf {\bibinfo {volume} {103}},\ \bibinfo
  {pages} {16008} (\bibinfo {year} {2021}{\natexlab{a}})}\BibitemShut {NoStop}%
\bibitem [{\citenamefont {Johnson}\ \emph {et~al.}(2011)\citenamefont
  {Johnson}, \citenamefont {Amin}, \citenamefont {Gildert}, \citenamefont
  {Lanting}, \citenamefont {Hamze}, \citenamefont {Dickson}, \citenamefont
  {Harris}, \citenamefont {Berkley}, \citenamefont {Johansson}, \citenamefont
  {Bunyk}, \citenamefont {Chapple}, \citenamefont {Enderud}, \citenamefont
  {Hilton}, \citenamefont {Karimi}, \citenamefont {Ladizinsky}, \citenamefont
  {Ladizinsky}, \citenamefont {Oh}, \citenamefont {Perminov}, \citenamefont
  {Rich}, \citenamefont {Thom}, \citenamefont {Tolkacheva}, \citenamefont
  {Truncik}, \citenamefont {Uchaikin}, \citenamefont {Wang}, \citenamefont
  {Wilson},\ and\ \citenamefont {Rose}}]{Johnson2011}%
  \BibitemOpen
  \bibfield  {author} {\bibinfo {author} {\bibfnamefont {M.~W.}\ \bibnamefont
  {Johnson}}, \bibinfo {author} {\bibfnamefont {M.~H.~S.}\ \bibnamefont
  {Amin}}, \bibinfo {author} {\bibfnamefont {S.}~\bibnamefont {Gildert}},
  \bibinfo {author} {\bibfnamefont {T.}~\bibnamefont {Lanting}}, \bibinfo
  {author} {\bibfnamefont {F.}~\bibnamefont {Hamze}}, \bibinfo {author}
  {\bibfnamefont {N.}~\bibnamefont {Dickson}}, \bibinfo {author} {\bibfnamefont
  {R.}~\bibnamefont {Harris}}, \bibinfo {author} {\bibfnamefont {A.~J.}\
  \bibnamefont {Berkley}}, \bibinfo {author} {\bibfnamefont {J.}~\bibnamefont
  {Johansson}}, \bibinfo {author} {\bibfnamefont {P.}~\bibnamefont {Bunyk}},
  \bibinfo {author} {\bibfnamefont {E.~M.}\ \bibnamefont {Chapple}}, \bibinfo
  {author} {\bibfnamefont {C.}~\bibnamefont {Enderud}}, \bibinfo {author}
  {\bibfnamefont {J.~P.}\ \bibnamefont {Hilton}}, \bibinfo {author}
  {\bibfnamefont {K.}~\bibnamefont {Karimi}}, \bibinfo {author} {\bibfnamefont
  {E.}~\bibnamefont {Ladizinsky}}, \bibinfo {author} {\bibfnamefont
  {N.}~\bibnamefont {Ladizinsky}}, \bibinfo {author} {\bibfnamefont
  {T.}~\bibnamefont {Oh}}, \bibinfo {author} {\bibfnamefont {I.}~\bibnamefont
  {Perminov}}, \bibinfo {author} {\bibfnamefont {C.}~\bibnamefont {Rich}},
  \bibinfo {author} {\bibfnamefont {M.~C.}\ \bibnamefont {Thom}}, \bibinfo
  {author} {\bibfnamefont {E.}~\bibnamefont {Tolkacheva}}, \bibinfo {author}
  {\bibfnamefont {C.~J.~S.}\ \bibnamefont {Truncik}}, \bibinfo {author}
  {\bibfnamefont {S.}~\bibnamefont {Uchaikin}}, \bibinfo {author}
  {\bibfnamefont {J.}~\bibnamefont {Wang}}, \bibinfo {author} {\bibfnamefont
  {B.}~\bibnamefont {Wilson}}, \ and\ \bibinfo {author} {\bibfnamefont
  {G.}~\bibnamefont {Rose}},\ }\href {\doibase 10.1038/nature10012} {\bibfield
  {journal} {\bibinfo  {journal} {Nature}\ }\textbf {\bibinfo {volume} {473}},\
  \bibinfo {pages} {194} (\bibinfo {year} {2011})}\BibitemShut {NoStop}%
\bibitem [{\citenamefont {Abel}\ \emph
  {et~al.}(2021{\natexlab{b}})\citenamefont {Abel}, \citenamefont {Blance},\
  and\ \citenamefont {Spannowsky}}]{Abel2021arxiv}%
  \BibitemOpen
  \bibfield  {author} {\bibinfo {author} {\bibfnamefont {S.}~\bibnamefont
  {Abel}}, \bibinfo {author} {\bibfnamefont {A.}~\bibnamefont {Blance}}, \ and\
  \bibinfo {author} {\bibfnamefont {M.}~\bibnamefont {Spannowsky}},\ }\href
  {https://arxiv.org/abs/2105.13945} {\bibfield  {journal} {\bibinfo  {journal}
  {arXiv:2105.13945}\ } (\bibinfo {year} {2021}{\natexlab{b}})}\BibitemShut
  {NoStop}%
\bibitem [{\citenamefont {Stella}\ \emph {et~al.}(2005)\citenamefont {Stella},
  \citenamefont {Santoro},\ and\ \citenamefont {Tosatti}}]{Stella05}%
  \BibitemOpen
  \bibfield  {author} {\bibinfo {author} {\bibfnamefont {L.}~\bibnamefont
  {Stella}}, \bibinfo {author} {\bibfnamefont {G.~E.}\ \bibnamefont {Santoro}},
  \ and\ \bibinfo {author} {\bibfnamefont {E.}~\bibnamefont {Tosatti}},\ }\href
  {\doibase 10.1103/PhysRevB.72.014303} {\bibfield  {journal} {\bibinfo
  {journal} {Phys. Rev. B}\ }\textbf {\bibinfo {volume} {72}},\ \bibinfo
  {pages} {014303} (\bibinfo {year} {2005})}\BibitemShut {NoStop}%
\bibitem [{\citenamefont {Shinomoto}\ and\ \citenamefont
  {Kabashima}(1991)}]{Shinomoto91}%
  \BibitemOpen
  \bibfield  {author} {\bibinfo {author} {\bibfnamefont {S.}~\bibnamefont
  {Shinomoto}}\ and\ \bibinfo {author} {\bibfnamefont {Y.}~\bibnamefont
  {Kabashima}},\ }\href {\doibase 10.1088/0305-4470/24/3/008} {\bibfield
  {journal} {\bibinfo  {journal} {J. Phys. A}\ }\textbf {\bibinfo {volume}
  {24}},\ \bibinfo {pages} {L141} (\bibinfo {year} {1991})}\BibitemShut
  {NoStop}%
\bibitem [{\citenamefont {Morita}(2007)}]{Morita07}%
  \BibitemOpen
  \bibfield  {author} {\bibinfo {author} {\bibfnamefont {S.}~\bibnamefont
  {Morita}},\ }\href {\doibase 10.1143/JPSJ.76.104001} {\bibfield  {journal}
  {\bibinfo  {journal} {J. Phys. Soc. Jpn.}\ }\textbf {\bibinfo {volume}
  {76}},\ \bibinfo {pages} {104001} (\bibinfo {year} {2007})}\BibitemShut
  {NoStop}%
\bibitem [{Note1()}]{Note1}%
  \BibitemOpen
  \bibinfo {note} {More concretely, we first determine the limit $x_{\protect
  \rm max}$ such that the ground state wavefunction at $m_i$ lies comfortably
  within the interval $[-x_{\protect \rm max},x_{\protect \rm max}]$. We then
  compute the energy eigenvalue $E_0$ using a grid spanning the interval with
  2048 grid points. The computed $E_0$ must remain unchanged when the number of
  grid points is decreased to 1024 and increased to 4096. With the converged
  grid at $m_i$, we repeat the procedure at $m_f$ to ensure that the same grid
  must guarantee the convergence of $E_0$ at the final mass as
  well.}\BibitemShut {Stop}%
\bibitem [{\citenamefont {Schmidt}\ and\ \citenamefont
  {Lorenz}(2017)}]{Schmidt17}%
  \BibitemOpen
  \bibfield  {author} {\bibinfo {author} {\bibfnamefont {B.}~\bibnamefont
  {Schmidt}}\ and\ \bibinfo {author} {\bibfnamefont {U.}~\bibnamefont
  {Lorenz}},\ }\href {\doibase https://doi.org/10.1016/j.cpc.2016.12.007}
  {\bibfield  {journal} {\bibinfo  {journal} {Comp. Phys. Commun.}\ }\textbf
  {\bibinfo {volume} {213}},\ \bibinfo {pages} {223} (\bibinfo {year}
  {2017})}\BibitemShut {NoStop}%
\bibitem [{\citenamefont {Schmidt}\ and\ \citenamefont
  {Hartmann}(2018)}]{Schmidt18}%
  \BibitemOpen
  \bibfield  {author} {\bibinfo {author} {\bibfnamefont {B.}~\bibnamefont
  {Schmidt}}\ and\ \bibinfo {author} {\bibfnamefont {C.}~\bibnamefont
  {Hartmann}},\ }\href {\doibase https://doi.org/10.1016/j.cpc.2018.02.022}
  {\bibfield  {journal} {\bibinfo  {journal} {Comp. Phys. Commun.}\ }\textbf
  {\bibinfo {volume} {228}},\ \bibinfo {pages} {229} (\bibinfo {year}
  {2018})}\BibitemShut {NoStop}%
\bibitem [{\citenamefont {Lorenz}\ and\ \citenamefont
  {Schmidt}(2020)}]{WavePacket}%
  \BibitemOpen
  \bibfield  {author} {\bibinfo {author} {\bibfnamefont {U.}~\bibnamefont
  {Lorenz}}\ and\ \bibinfo {author} {\bibfnamefont {B.}~\bibnamefont
  {Schmidt}},\ }\href@noop {} {\enquote {\bibinfo {title} {{WavePacket
  (C++/Python): Time-dependent simulation of open and closed quantum systems.
  Version 0.3.1.}}}\ } (\bibinfo {year} {2020}),\ \bibinfo {note}
  {\url{https://sourceforge.net/projects/wavepacket/}}\BibitemShut {NoStop}%
\bibitem [{Note2()}]{Note2}%
  \BibitemOpen
  \bibinfo {note} {Convergence of trajectories with respect to $dt$ is checked
  by halving it (to 0.05) and then seeing that recalculated results remain
  invariant.}\BibitemShut {Stop}%
\bibitem [{Note3()}]{Note3}%
  \BibitemOpen
  \bibinfo {note} {Due to the periodic oscillatory behavior of the data points,
  instead of peforming a least square fit, we have chosen instead to manually
  adjust the vertical displacement of the line $T^{-2}$ to fit those data
  points at the crests of the oscillatory curve. We find that this reflects
  more accurately the observed gradients of the curves.}\BibitemShut {Stop}%
\bibitem [{\citenamefont {Suzuki}\ and\ \citenamefont
  {Okada}(2005)}]{Suzuki2005}%
  \BibitemOpen
  \bibfield  {author} {\bibinfo {author} {\bibfnamefont {S.}~\bibnamefont
  {Suzuki}}\ and\ \bibinfo {author} {\bibfnamefont {M.}~\bibnamefont {Okada}},\
  }\href {\doibase 10.1143/JPSJ.74.1649} {\bibfield  {journal} {\bibinfo
  {journal} {J. Phys. Soc. Jpn.}\ }\textbf {\bibinfo {volume} {74}},\ \bibinfo
  {pages} {1649} (\bibinfo {year} {2005})}\BibitemShut {NoStop}%
\bibitem [{Note4()}]{Note4}%
  \BibitemOpen
  \bibinfo {note} {For the plane wave $e^{ikx}$, one has $j(x,t)=\protect \frac
  {\hbar k}{m}=\protect \frac {p}{m}=v$, which is simply the velocity of the
  particle. Hence, the probability current can be interpreted as a
  generalization of the velocity field, indicating both the speed and direction
  of motion of the particle.}\BibitemShut {Stop}%
\bibitem [{Note5()}]{Note5}%
  \BibitemOpen
  \bibinfo {note} {The simulated annealing calculations in Secs. \ref
  {sec.SA.log schedule} and \ref {sec.SA.linear schedules} are performed with
  $dt=0.1$, to be consistent with the time step used in quantum
  annealing.}\BibitemShut {Stop}%
\end{thebibliography}
